\documentclass[12pt,letterpaper]{article}
\usepackage[margin=1in]{geometry}

\usepackage{amsmath}
\usepackage{amsfonts}
\usepackage{amsthm}
\usepackage{amssymb}
\usepackage{graphicx}
\usepackage{url}
\usepackage[authoryear]{natbib}
\usepackage[caption=false]{subfig}
\usepackage{enumitem}

\title{Co-design and Co-simulation for Engineering Systems: Insights from the Sustainable Infrastructure Planning Game}

\author{Paul T. Grogan  (\url{pgrogan@stevens.edu}) \\ School of Systems and Enterprises \\ Stevens Institute of Technology \\ Hoboken, NJ, USA}

\begin{document}

\maketitle

\begin{abstract}
This paper draws on perspectives from co-design as an integrative and collaborative design activity and co-simulation as a supporting information system to advance engineering design methods for problems of societal significance. Design and implementation of the Sustainable Infrastructure Planning Game provides a prototypical co-design artifact that leverages the High Level Architecture co-simulation standard. Three role players create a strategic infrastructure plan for agriculture, water, and energy sectors to meet sustainability objectives for a growing and urbaninzing population in a fictional desert nation. An observational study conducts 15 co-design sessions to understand underlying dynamics between actors and how co-simulation capabilities influence design outcomes. Results characterize the dependencies and conflicts between player roles based on technical exchange of resource flows, identifying tension between agriculture and water roles based on water demands for irrigation. Analysis shows a correlation between data exchange, facilitated by synchronous co-simulation, and highly-ranked achievement of joint sustainability outcomes.  Conclusions reflect on the opportunities and challenges presented by co-simulation in co-design settings to address engineering systems problems.

\end{abstract}

\section{Introduction}

Pursuit of societally-relevant objectives such as resource security or sustainability presents a challenge for traditional systems engineering and design methods because no single actor has complete knowledge of or control over all constituent systems. For example, consider the link between water and energy resources in infrastructure: despite energy-intensive processes for water desalination, distribution, and treatment and water-intensive processes for energy extraction, refining, and cooling, energy and water policies are largely developed independent of each other \citep{Hussey_2012_EnergyWaterNexus}. Further interdependencies with other resources such as agriculture and food pose multi-level coordination challenges \citep{PahlWostl_2019_GovernanceWaterEnergyFood}. Myopic decisions based on incomplete or inaccessible information can create significant and lasting harm to natural resources without malice or intent, leading to situations such as the current global groundwater crisis \citep{Famiglietti_2014_GlobalGroundwaterCrisis}.

Given inherent human limits to knowledge aggregation and centralization of control, \textit{co-design} frames design as a social process of ``joint inquiry and imagination'' by integrating diverse viewpoints \citep{Steen_2013_CoDesignProcess}. Intertwined with related concepts such as participatory design, concurrent engineering, and collaborative design, co-design places information exchange at the heart of problem exploration, definition, and perception, well before conceiving of and evaluating potential solutions.

Information systems (IS) facilitate interaction and information exchange between design actors in engineering design \citep{McMahon_2004_KnowledgeManagement}. While internet-enabled IS have long been envisioned as a platform for co-design \citep{Li_2004_InternetEnabledSystem}, comparatively few engineering design activities leverage IS for innovative design processes today. For example, model-based systems engineering (MBSE) is perhaps the most widespread IS-enabled design process but is still in its early phases and runs largely parallel to traditional document-driven systems engineering \citep{Madni_2018_MBSE}. 

Going beyond model exchange in MBSE, co-simulation is a modeling technique for dynamic information exchange that leverages distributed IS to study a joint problem by composing constituent parts \citep{Gomes_2018_CoSimulation}. Co-simulation provides a technical foundation for co-design activities on which participating organizations explore and define the problem from different perspectives. However, beyond modest adoption in defense, automative, and aerospace domains, co-simulation remains a novel technique lacking in supporting methods and processes to support engineering systems design activities at large scales. At the same time, existing gaming applications that blend participatory co-design and simulation activities do not leverage co-simulation techniques.

This paper discusses the design and evaluation of the Sustainable Infrastructure Planning Game (SIPG)\footnote{SIPG is available under an open source license at \url{https://github.com/code-lab-org/sipg} but currently requires an HLA runtime infrastructure (RTI) to operate.} as a co-simulation artifact to support co-design studies \citep{Grogan_2014_InteroperableSimulationGaming}. While based on generalizable constructs, SIPG formulates a strategic infrastructure planning scenario for resource security and sustainability goals with three role-players that control agriculture, water, and energy systems. SIPG provides a platform on which to conduct and observe co-design sessions and understand how co-simulation influences co-design activities. 

This paper connects initial results of SIPG design sessions reported in \cite{Grogan_2016_CollaborativeDesignSIPG} with broader discussion of co-design and co-simulation. Following a design science perspective \citep{Hevner_2004_DesignScience}, this work presents SIPG as a prototypical IS artifact that provides dynamic information exchange during co-design settings. A co-simulation modeling framework implemented by the the High Level Architecture (HLA) standard enforces spatial-temporal resource interfaces while enabling decentralized control over role-specific simulations. Artifact evaluation collects data from an observational study of 15 co-design sessions to understand scenario-specific game dynamics and how information exchange enabled by co-simulation influences process and outcome variables across session variants with strong and weak adoption of co-simulation.

The remainder of this paper is structured as follows. Section 2 reviews background literature on co-design and co-simulation to refine research objectives. Section 3 describes the formulation and technical implementation of the SIPG co-simulation application. Section 4 constructs an observational study to investigate the SIPG game dynamics and how co-simulation influences collective outcomes. Section 5 presents results, statistical analysis, and discussion. Finally, Section 6 concludes by revisiting the role of co-simulation in co-design.

\section{Background Literature} \label{sec:background}

\subsection{Perspectives on Co-design}

Co-design encompasses design activities and processes that generally exchange information across design roles. Similar to the broader topic of integrated assessment in environmental and sustainability literature \citep{Rotmans_1998_MethodsIA}, co-design has both analytical and participatory methods. From a technical perspective, analytical methods perform model, scenario, and risk analyses to represent and structure scientific knowledge. From a social perspective, participatory methods such as expert panels, Delphi methods, gaming, policy exercises, and focus groups draw on social sciences to involve a broad stakeholder set.

Technical co-design in literature refers to simultaneous decision-making across traditionally-sequential disciplines enabled by a shared model (i.e.~variables, objectives, and constraints) as a type of multi-disciplinary design optimization (MDO) \citep{Allison_2014_CoDesign,Azad_2020_RobustCoDesign}. Tighter coupling between decisions allows knowledge in one domain to more directly influence another without unnecessary constraints or delay from multiple iterations, resulting in a more desirable solution.

Concurrent engineering (CE) shares a technical perspective with co-design as an integrated method to coordinate design activities to achieve holistic objectives but also includes disciplinary human actors inside the system boundary \citep{Holt_2010_IntegratedApproachDFX}. CE broadly encompasses the team (people), model (shared knowledge), tools, process, and facility for a design activity \citep{Knoll_2018_ReviewCE}. Alternative CE strategies seek either to decouple tasks to reduce time or increase coupling between tasks to improve quality \citep{Eppinger_1991_ModelBasedConcurrentEngineering}. While there is generally no centralized optimizer as in MDO, CE shares a common understanding of the problem and objectives and relies on integrating roles such as systems engineering to facilitate activities.

In contrast to technical solution processes in MDO and CE, other perspectives view co-design as a negotiated solution process between different viewpoints \citep{Detienne_2005_ViewpointsCoDesign}. Social processes of imaginative creativity and mutual knowledge exchange build on more than 40 years of participatory design that link the designer with other design actors including customers \citep{Sanders_2008_CoCreation} to broaden the set of stakeholders who influence design decisions \citep{Carroll_2007_ParticipatoryDesign}. Framed as a ``process of joint inquiry and imagination,'' co-design connects individual practices, experiences, and knowledge with collective communication, cooperation, and change \citep{Steen_2013_CoDesignProcess}. Co-design activities seek to overcome barriers to shared understanding at individual, project, and organizational levels \citep{Kleinsmann_2008_BarriersEnablersCoDesign}.

Collaboration is at the heart of co-design activities; however, collaborative design activities can succumb to complexity if poorly structured \citep{Suh_2009_DesigningEngineeringCollaboration}. Research on collaborative engineering applies results from organization science, social cognition, social choice, and decision science to engineering practice to work towards a common goal with limited resources or conflicting interests \citep{Lu_2007_CollaborativeEngineering}. Proposed methods for engineering collaboration via negotiation outline a four-step process to manage social interactions, construct shared understanding, discourse preferences, and finally attain agreement \citep{Lu_2007_CollaborativeEngineering}. Mediation may benefit negotiation processes to achieve joint decisions among two or more parties while maximizing social welfare \citep{Klein_2003_NegotiatingComplexContracts}.

\subsection{Information Systems and Co-simulation}

Information systems (IS) are artifacts that extend human cognitive limits, exchange information, record mental efforts, and mediate critique and negotiation in design settings \citep{Arias_2000_TranscendingHumanMind}. Research on computer-supported collaborative design combines broader fields of computer supported cooperative work and human-computer interaction to study the use of computers in design activities \citep{Shen_2008_CSCD}. Typical computer support functions include visualization, cross-disciplinary information exchange, and integrated lifecycle analysis \citep{Li_2006_CollaborativeDevelopment} and common challenges include interoperability, integration, facilitation, and change management \citep{Shen_2010_SystemsIntegrationCollaboration}.

Model-driven or model-centric design activities leverage IS to capture and exchange disciplinary knowledge across team members through a shared system model \citep{Ramos_2012_MBSE}. Model creation helps facilitate learning by eliciting and formalizing knowledge, synthesizing new feedback loops to support or refute hypotheses, and sharpening scientific (versus position-driven) solution skills \citep{Sterman_1994_LearningComplexSystems,Venix_1999_GroupModelBuilding}. However, computer-based models can also pose barriers due to limited acceptance, insufficient time to complete a feedback loop, poor user-friendliness, high model complexity, and inflexibility to incorporate issues of interest \citep{deKraker_2011_ParticipatoryIntegratedAssessment}. 

Alternative IS architectures balance centralized and distributed control over the design activity \citep{Whitfield_2002_DistributedDesignCoordination}. Centralized control exerts strict requirements on modeling languages, interfaces, or even model co-location which permits more efficient or effective solution processes (i.e.,~MDO) but presents practical challenges in participatory settings, especially across organizational boundaries where cultural or even legal issues may limit ceding of control. Distributed control schemes push integration requirements to more abstract layers which brings additional challenges of higher network and processing requirements and added overall complexity. 

Long-running efforts dating to the early phases of CE seek to improve distributed design by encapsulating, rather than standardizing or unifying, tool data and models \citep{Cutkosky_1993_PACT}. The broader topic of model interoperability describes the ability of ``multiple separate entities to interact, collaborate, or utilize each other to achieve higher level goal or their own goals'' and can be applied at multiple levels ranging from technical interconnectivity to programmatic coordination \citep{Mordecai_2016_ModelBasedInteroperability}. Increasing degrees of model interoperability enforce common information exchange protocols, syntax, semantics for static interactions and shared knowledge of methods, state changes, and overall assumptions for dynamic interactions \citep{Tolk_2007_LCIM}.

Co-simulation is a technique to couple the execution of multiple simulators to facilitate dynamic information exchange across disciplines, domains, or organizations \citep{Gomes_2018_CoSimulation}. Co-simulation methods range from acausal continuous time modeling languages which align constituent models through dynamic equations \citep{Mattsson_1998_SystemModelingModelica} to general discrete event frameworks which build on parallel and distributed simulation to synchronize state and maintain causality across multiple logical processes \citep{Fujimoto_2000_PADS}. Current standards include the Functional Mockup Interface (FMI) \citep{Modelica_2019_FMI} for continuous time simulation and High Level Architecture (HLA) \citep{IEEE_2010_HLA} for discrete event simulation.

Simulation-based methods for infrastructure systems still face difficulties with sharing models and data across organizational boundaries, considering both hard and soft infrastructure, exchanging mutual dependencies, and validation for novel or unexpected scenarios \citep{Ouyang_2014_ReviewModelingSimulation}. Co-simulation must overcome differing time-scales and resolutions or fidelity of component models \citep{Pederson_2006_CriticalInfrastructure}. Each domain carries different assumptions, data dependencies, and numerical requirements for time step sizes, scaling limits, or computational algorithms that generally limit the adoption of existing domain-specific models \citep{Rinaldi_2001_CriticalInfrastructure}. Application of standards like HLA for co-simulation has thus far have been limited due, in part, to industry focus on inexpensive, limited, and disposable models using commercial off the shelf packages compared to the relatively expensive runtime infrastructure (RTI) licenses, general-purpose programming language, high complexity, and limited community of experts for HLA \citep{Boer_2009_DistributedSimulationIndustry}. Alternative options study simpler service-oriented architectures for infrastructure modeling with centralized event processing and significantly reduced functionality \citep{Tolone_2008_CriticalInfrastructure}.

\subsection{Simulation Gaming}

Applications of simulation to co-design problems where participants role-play decision-making actors in an interactive design session can be described as simulation gaming or, simply, gaming \citep{Grogan_2017_GamingMethods}. Simulation emphasizes technical system behavior that can be represented with a computational model while gaming emphasizes distinctly human and social behavior such as cognitive bias, bounded rationality, culture, politics, strategy, ethics, and morality. Compared to a static dependencies in engineering co-design, simulation gaming exchanges dynamic dependencies over a simulated timeline. Repeated interactions between role players contribute to complex interdependencies and conflicts. Partly an exploratory device and partly an experimental platform, games have been applied over a substantial history to study a wide range of collective decision-making problems spanning military tactics, supply chain logistics, international crises, and urban planning  \citep{Mayer_2009_GamingOfPolicy}.

A series of ``infra-games'' developed over the past two decades apply gaming methods to study infrastructure planning problems. While each game focuses on a different problem, common features emphasize collaborative decision-making and strategic behaviors. For example:
\begin{enumerate}
    \item The Urban Network Game seeks insights to opportunities and threats to developing urban networks of cities with good transportation connectivity \citep{Mayer_2004_UrbanNetwork},
    \item Infrastratego studies strategic behavior in a liberalizing Dutch electricity market and effectiveness of different regulatory regimes \citep{Kuit_2005_Infrastratego},
    \item SprintCity studies the interrelations between rail infrastructure and urban development near stations \citep{Nefs_2010_SprintCity,Mayer_2010_SprintCity}, and
    \item  SimPort-MV2 demonstrates complexities of a large land reclamation project at the Port of Rotterdam in the Netherlands \citep{Bekebrede_2015_UnderstandingComplexity}.
\end{enumerate}
Each game uses simulation to a different degree to model technical infrastructure systems but all focus on the participatory decisions to either generate insights about a domain-specific problem (e.g., strategic behaviors in electricity markets for Infrastratego) or develop generalizable knowledge for a class of problems (e.g., managing complex infrastructure for SimPort-MV2). Usually structured as an interactive simulation, role players input decisions to a simulation model which computes and disseminates results throughout a dynamic scenario. Not all games require high-tech IS to achieve simulation objectives and using simple physical props such as sponges to represent train positions can facilitate rapid system development and prototyping \citep{Meijer_2015_PowerOfSponges}.

While a body of literature addresses design of simulation games as ``design-in-the-small'' \citep[see][]{Klabbers_2003_Introduction}, no comparable literature considers co-simulation because, in most cases, a centralized IS satisfies all research goals and is easier to implement. Research principals, rather than participants, develop technical simulations for games like SimPort-MV2 and Infrastratego to focus studies on collaborative processes. Gaming applications of co-simulation only appear in domains with a precedent for co-simulation in practice such as defense and emergency response \citep{Prasithsangaree_2004_UTSAF,McLean_2008_IncidentManagement}.

\subsection{Research Objectives}

\begin{figure}
    \centering
    \includegraphics[scale=1]{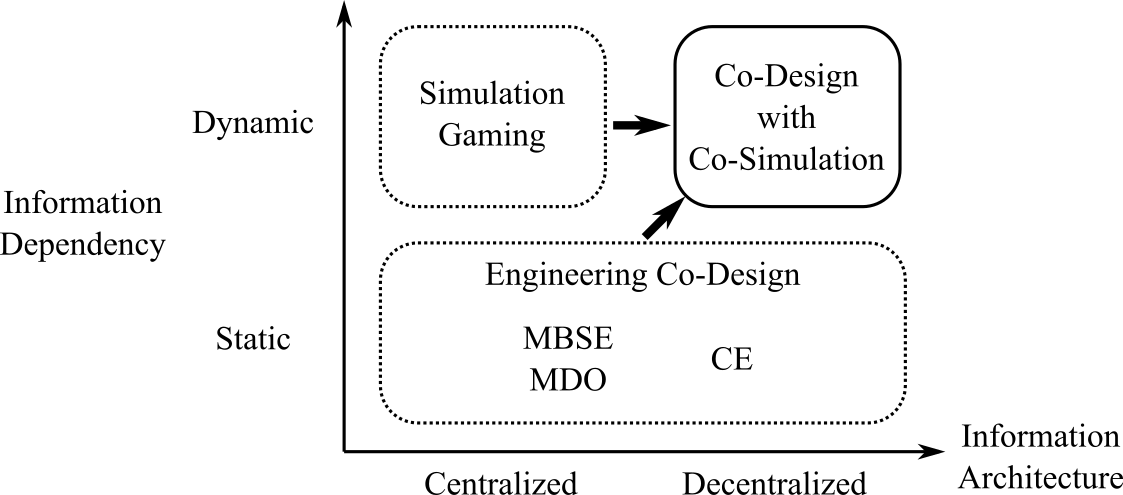}
    \caption{Co-design with co-simulation supports dynamic information dependency with decentralized control over constituent information system components (models).}
    \label{fig:architecture}
\end{figure}

To synthesize preceding sections, Fig.~\ref{fig:architecture} illustrates the relationship between engineering co-design, simulation gaming, and co-simulation. Engineering co-design uses technical solution processes like MDO, MBSE, and CE to solve a design problem by communicating static dependencies between design actors. Centralized IS architectures integrate all information in a single model while a decentralized IS provide distributed control over constituent models. Simulation gaming draws on a broader set of negotiated solution processes to exchange dynamic dependencies between actors in a game session; however, existing gaming methods use centralized IS contributed by a principal rather than participants. Co-design with co-simulation seeks to exchange dynamic information dependencies within a collaborative process (like simulation gaming) while providing decentralized control over constituent models.

Co-design with co-simulation spans both technical integration and social collaboration to build shared understanding of dynamic dependencies in large-scale engineering systems like infrastructure systems where there may be significant barriers to centralize technical components. However, there remains a gap to understand how co-simulation contributes to technical and negotiated solution activities. Within the context of a representative infrastructure planning scenario, this paper addresses the top-level research question:
\begin{itemize}[label={}]
    \item How can co-simulation artifacts technically integrate and provide dynamic information exchange among design actors during co-design activities?
\end{itemize}
The response follows a design science research methodology \citep{Hevner_2004_DesignScience} to create and evaluate the utility of a co-simulation IS artifact in a co-design setting. In other words, the equivalent research hypothesis is that co-simulation provides utility for co-design settings through dynamic information exchange.

The following sections develop a representative co-design scenario based on a 30-year strategic infrastructure planning activity for a fictional desert nation. Three role-players control infrastructure systems with dynamic resource dependencies while pursuing individual and joint sustainability objectives. Technical details of a prototype co-simulation artifact explain how the HLA standard integrates constituent system models with decentralized control. Finally, an observational human subjects study evaluates the co-simulation artifact for co-design by investigating what game dynamics emerge between design actors and analyzing how co-simulation features influence design processes and outcomes.

\section{Sustainable Infrastructure Planning Game} \label{sec:application}

This section discusses the design and implementation of the Sustainable Infrastructure Planning Game (SIPG) as a co-design scenario and co-simulation artifact that uses the HLA standard \citep{IEEE_2010_HLA} to technically integrate constituent infrastructure system simulations. Although the underlying concepts are generalizable to any infrastructure system, SIPG draws on a specific design scenario for a fictional desert nation loosely based on contextual features of Saudi Arabia between 1950--2010. It defines three player roles who exert control over agriculture, water, and energy infrastructure systems with objectives based on multi-dimensional system attributes. Some objectives are aligned towards collective sustainability objectives while others are in conflict between roles.

\subsection{Co-design Scenario}

\begin{figure}
    \centering
    \includegraphics{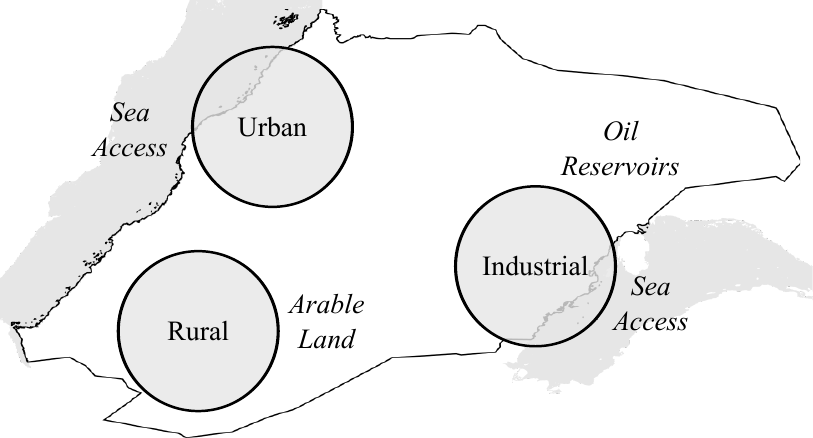}
    \caption{Urban, Rural, and Industrial regions of a fictional desert nation.}
    \label{fig:framework}
\end{figure}

The SIPG co-design scenario is a strategic planning exercise to create a 30-year infrastructure development plan for Idas Abara, a fictional desert nation with a petroleum-based economy. The scenario takes place in the year 1980, as infrastructure pressures mount from resource demands of rapid population growth and urbanization. Urban, industrial and rural geographic regions illustrated in Fig.~\ref{fig:framework} aggregate infrastructure, each with unique population dynamics and suitability for new infrastructure projects. The urban and industrial regions can access seawater for desalination, the industrial region has vast (but finite) oil reservoirs, and the rural region has plentiful arable land.

Viewing the scenario from the driver-pressure-state-impact-response (DPSIR) framework \citep{Tscherning_2012_DPSIR}, driving forces are linked to a rapidly growing and urbanizing population. In addition to annual population growth rates exceeding 3\%, urban lifestyles increase per-capita demands for food, water, oil, and, most significantly, electricity (e.g.~for air conditioning). Environmental pressures include withdrawals from non-renewable ``fossil'' water aquifers and oil reservoirs and increased emissions. The potential impacts of environment changes are wide-reaching and significant---depletion of water resources has dire consequences for society (broadly) but also limits efforts to diversify the economy through agriculture. Reduction of oil resources from reservoir depletion can also trigger a financial crisis, as oil exports currently sustain the national economy. While only a small piece of sustainability, the response considered in this scenario develops a strategic infrastructure plan to provide necessary resources while sustaining economic and environmental conditions. 

The design scenario includes player roles for agriculture, water, and energy (oil and electricity) sectors and a non-player role for all other societal activities such as commercial and residential demands. Players choose when and where to construct and operate new infrastructure elements to transform or transport resources to meet societal demands. Each infrastructure element consumes capital expenses during construction and operational expenses during its lifecycle. A nation-wide budget limit establishes a soft constraint on the capital expenses allowed each year. Plans that exceed the annual budget limit are permitted with notation of the over-budgeted period(s).

Co-design activities identify new capital projects to supplement existing infrastructure and simulate resource production and distribution over the 30-year planning period. Simulation outputs quantify several metrics to inform decision-making. Iterative design and evaluation helps to uncover coordination challenges in pursuit of four equally-weighted joint sustainability objectives:
\begin{enumerate}
    \item Food security as the fraction of demands satisfied by domestic production,
    \item Water security as the expected aquifer lifetime at current withdrawal rates,
    \item Oil security as the expected reservoir lifetime at current withdrawal rates, and
    \item Financial security as the net revenue of all infrastructure systems.
\end{enumerate}

Design conflicts arise from three linked sources. First, interest to strengthen food security by increasing domestic food production puts additional demands for irrigation, greatly diminishing available water resources in non-renewable aquifers. Subsequent efforts to increase desalination capacity greatly amplify pressures on power generation and domestic oil consumption, diminishing revenue from profitable oil export. Finally, efforts to increase renewable power generation require large capital expenses that compete with desalination projects for limited budget capacity.

\subsection{Co-simulation Platform and Interfaces}

This section formulates a co-simulation IS artifact to technically integrate role-specific constituent models and provide dynamic information exchange for the SIPG design scenario. Co-simulation requires an interface between each pair of dependent systems. To simplify the co-simulation architecture, this section adopts the infrastructure system-of-systems (ISoS) modeling framework \citep{Grogan_2015_ModelingFrameworkISOS}  as a common interface among all constituent systems that can be implemented using the HLA \citep{Grogan_2018_InfrastructureSystemSimulation}. The ISoS framework defines contextual, structural, and behavioral templates to guide constituent model development efforts. The interoperability interface defines a service contract for necessary start-up, synchronization, and shut-down procedures to model resource exchanges across system boundaries.

The contextual template defines application-specific constructs for spatial and temporal boundaries. Nodes define geographic units of aggregation where resources can be freely exchanged. SIPG nodes represent the three regions (urban, industrial, and rural). The time advancement strategy defines a common time step duration with several iteration periods to resolve dependencies. SIPG uses an aggregated one-year time step with four iterative periods. Finally, a set of resource types describe the substances exchanged between systems. Key SIPG resources include water, electricity, oil, food, and currency.

Structural templates define infrastructure elements as resource-conveying edges at or between nodes. Production and storage elements have the same origin and destination while distribution elements have different origins and destinations. Elements assume a lifecycle state that transitions between five sequential phases: empty (pre-initiation), commissioning, operating, decommissioning, and null (post-termination). Additional state variables set, for example, operational production, withdrawal, or distribution levels during each time period.

Behavioral templates express four key infrastructure resource functions. Storing or retrieving functions add or remove resources from a stored stock or buffer. Transforming functions convert input resources at the source node to output resources at the destination node. Transporting functions move input resources at the source node to the destination node. Exchanging functions transfer resources from an origin in one system to a destination in another.

Co-simulation only disseminates resource exchanging behaviors across system boundaries. The HLA requires a shared federation object model (FOM) to describe the syntax and semantics of exchanged information and a federation agreement to document required activities during a simulation execution. As a component of the FOM, Table~\ref{tab:fom} describes key object attributes for constituent system elements (all inheriting from a base Element class). All attributes are aggregated to annual temporal scales and nodal spatial scales. Ideally, supply should meet demand for each resource type, e.g.~Food Out from the agriculture system should equal Food In for the co-located Societal System; however, an iterative convergence process driven by system controllers to optimize production under interdependency contributes small errors. \cite{Grogan_2018_InfrastructureSystemSimulation} provides more details on the HLA implementation for interested readers.

\begin{table*}
    \small
    \caption{Key object attributes to exchange during co-simulation.}
    \begin{tabular}{llll}
        \hline
        Class & Attribute & Data Type & Semantics \\
        \hline
        Generic Element & Name & String & Unique element identifier \\
        & Location & String & Node location identifier \\
        & Currency Flow & Float & Net annual cash flow ($\S$) \\
        & Capital Expenses & Float & Annual capital expenses ($\S$) \\
        
        \quad Agriculture System & Water In & Float & Annual demand (MCM) \\
        & Food Out (Societal) & Float & Annual supply (GJ) \\
        
        \quad Water System & Electricity In & Float & Annual demand (TWh) \\
        & Water Out (Agriculture) & Float & Annual supply (MCM) \\
        & Water Out (Societal) & Float & Annual supply (MCM) \\
        
        \quad Petroleum System & Electricity In & Float & Annual demand (TWh) \\
        & Oil Out (Societal) & Float & Annual supply (Mtoe) \\
        & Oil Out (Electrical) & Float & Annual supply (Mtoe) \\
        
        \quad Electrical System & Oil In & Float & Annual demand (Mtoe) \\
        & Electricity Out (Water) & Float & Annual supply (TWh) \\
        & Electricity Out (Societal) & Float & Annual supply (TWh) \\
        
        \quad Societal System & Water In & Float & Annual demand (MCM) \\
        & Food In & Float & Annual demand (GJ) \\
        & Oil In & Float & Annual demand (Mtoe) \\
        & Electricity In & Float & Annual demand (TWh) \\
        \hline
        \multicolumn{4}{l}{Units: $\S$: fictional currency; GJ: gigajoule; MCM: million cubic meters; } \\
        \multicolumn{4}{l}{Mtoe: million tons oil equivalent; TWh: terawatt hour} \\
    \end{tabular}
    \label{tab:fom}
\end{table*}

\begin{figure}
    \centering
    \includegraphics[width=3.25in]{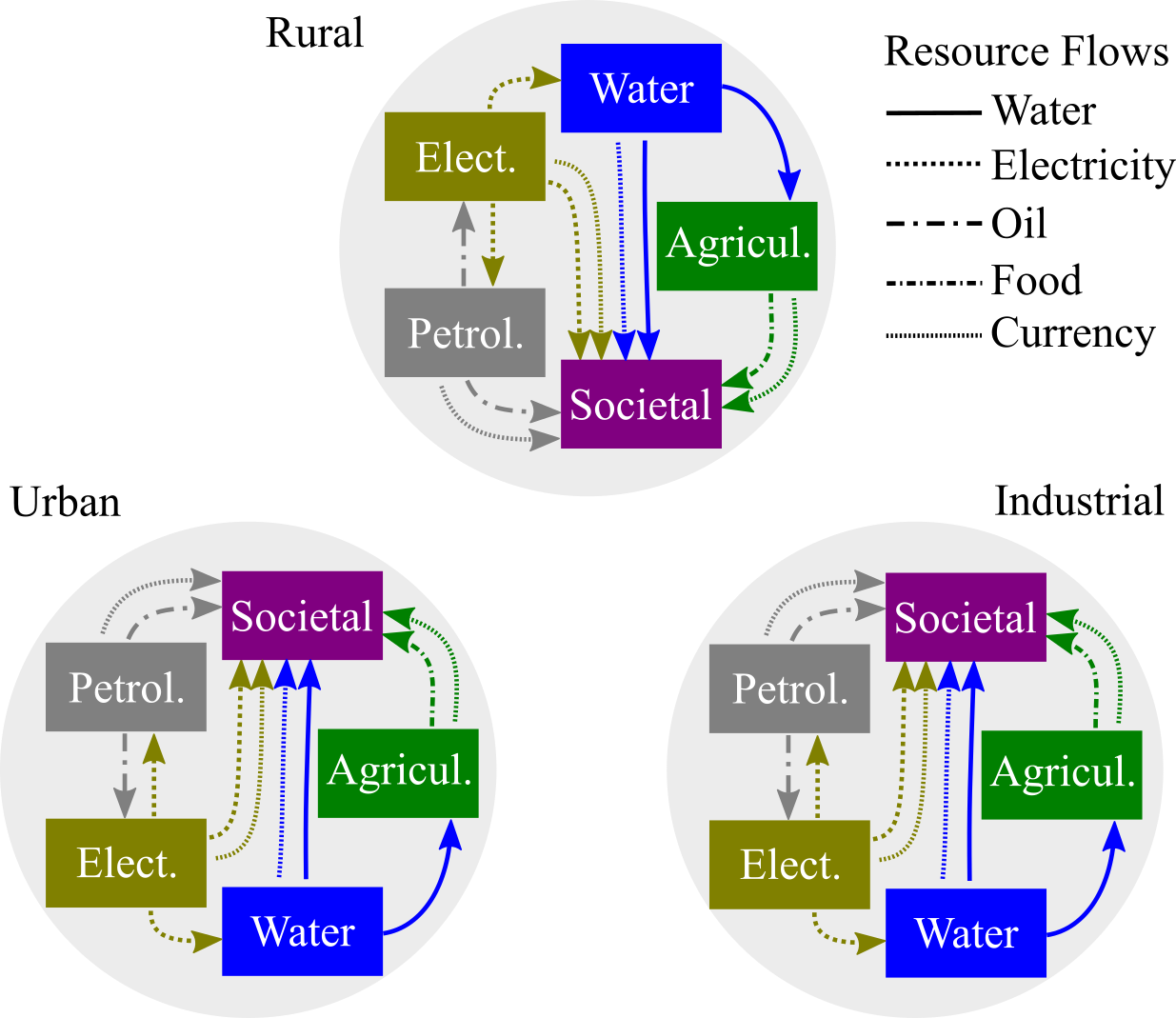}
    \caption{Constituent infrastructure systems at each node dynamically exchange directed resources using interfaces defined by the co-simulation platform.}
    \label{fig:sosModel}
\end{figure}

The resulting system-of-systems model in Fig.~\ref{fig:sosModel} includes societal, agricultural, water, and energy (electrical and petroleum) infrastructure systems at each node. Within each system, production elements transform raw to refined resources and distribution elements move resources between nodes. However, at the co-simulation level, dynamic interactions are characterized only by resource exchanges. Key resource flows supply water to the agricultural system for irrigation, electricity to the water and petroleum systems to power pumps, wells, and desalination plants, and petroleum to the electricity system for thermal generation. The societal system consumes food, water, electricity, and oil to satisfy demands and stores the net balance of currency.

\subsection{Constituent Model Implementation Overview}

The co-simulation interface defines directed resource exchanges between constituent models but not their internal state or behaviors which is private information for each role player. This section provides a brief overview of the objectives and available infrastructure models for agriculture, water, and energy roles as examples of sensitive and domain-specific information that may not be shareable on a centralized IS platform. See Appendix A for detailed documentation of infrastructure models and Appendix B for formulation of objective metrics.

The agriculture model controls land allocation for food production and roads to transport food between regions. Role-specific objectives are:
\begin{enumerate}
    \item Food security as the fraction of demands satisfied by domestic production,
    \item Financial security as the net revenue of the agriculture system, and
    \item Political power as the total capital allocated to the agriculture system.
\end{enumerate}
Available infrastructure elements include small and large fields for production and small and large roads for distribution which are controlled each time step to meet demands at minimum cost. Food production is constrained by arable land area and available workers as a fraction of population and requires requires water for irrigation. Regions export surplus food for a profit and import to meet deficits.

The water models controls desalination plants and non-renewable (i.e.~``fossil'') aquifer stocks. Role-specific objectives are:
\begin{enumerate}
    \item Aquifer security as the expected lifetime at current withdrawal rates,
    \item Financial security as the net revenue of the water system, and
    \item Political power as the total capital allocated to the water system.
\end{enumerate}
Available infrastructure elements include small, large, and huge desalination plants which are controlled each time step to meet demands at minimum cost. Deficits in desalination supply require regions to lift water from aquifers and, when depleted, import at great expense. Both desalination and lifting require electricity. No transport of water is permitted between regions due to large pumping expenses. 

The energy role composes both petroleum systems (oil wells and pipelines) and electrical systems (power plants) which are tightly coupled because oil pumping requires electricity and thermal power generation requires oil as feed stock. Role-specific objectives are:
\begin{enumerate}
    \item Reservoir security as the expected lifetime at current withdrawal rates,
    \item Financial security as the net revenue of the energy system, and
    \item Political power as the total capital allocated to the energy system.
\end{enumerate}
Available petroleum infrastructure include small and large well pumps for production and small and large pipelines for distribution. Available electricity infrastructure for generation include small and large thermal plants and small and large solar plants. All infrastructure elements are controlled at each time step to meet demands at minimum cost. Regions export surplus oil for profit, import oil to meet supply deficits, and use low-efficiency ``private'' thermal generation to meet electricity deficits.

\subsection{Graphical User Interface}

A graphical user interface allows design actors to modify role-specific infrastructure systems, execute a co-simulation, and view outputs. Inputs define a sector-specific infrastructure plan composing the type, location, and time to build each new element. Outputs visualize key resource flows and quantify figures of merit associated with role objectives.

\begin{figure}
    \begin{minipage}{.39\linewidth}
        \centering
    	\subfloat[Simulation State and Control]{%
    		\includegraphics[scale=0.5,trim={1px 14px 724px 40px},clip]{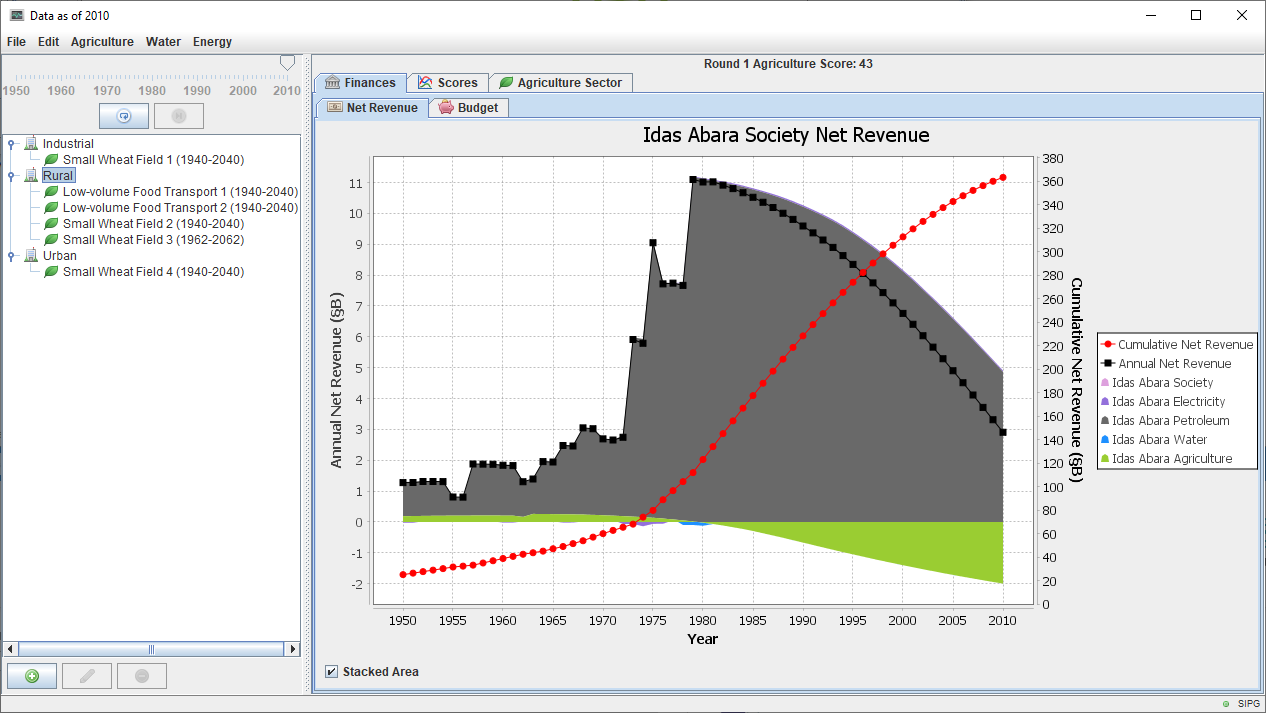}
    		\label{fig:gui_input_panel}}
	\end{minipage}
    \begin{minipage}{.49\linewidth}
        \centering
    	\subfloat[Infrastructure Template]{%
    		\includegraphics[scale=0.5]{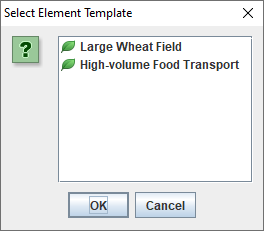}
    		\label{fig:gui_input_menu}}
    	
    	\subfloat[Infrastructure Element Editor]{%
    		\includegraphics[scale=0.5]{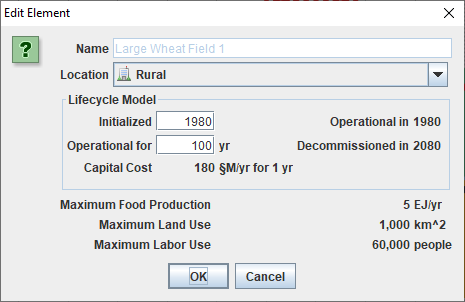}
    		\label{fig:gui_input_dialog}}
	\end{minipage}
    \caption{Input GUI components control simulation execution and edit infrastructure.}
    \label{fig:gui_input}
\end{figure}

The input panel includes simulation controls and a list of existing elements. Simulation controls initialize and run a co-simulation execution. Figure~\ref{fig:gui_input_panel} shows the current set of infrastructure elements, grouped by location. Players can add or edit new elements, choosing from a role-specific menu of templates in Fig.~\ref{fig:gui_input_menu}. Each infrastructure element displays key lifecycle and operational information shown in Fig.~\ref{fig:gui_input_dialog} such as capital cost, lifespan, and resource production.

\begin{figure*}
	\subfloat[Societal Net Revenue]{%
		\includegraphics[width=0.48\linewidth,trim={236px 17px 5px 53px},clip]{figures/panel_revenue.png}
		\label{fig:gui_net_revenue}}
	\hfill
	\subfloat[Capital Expense Budget]{%
		\includegraphics[width=0.48\linewidth,trim={236px 17px 5px 53px},clip]{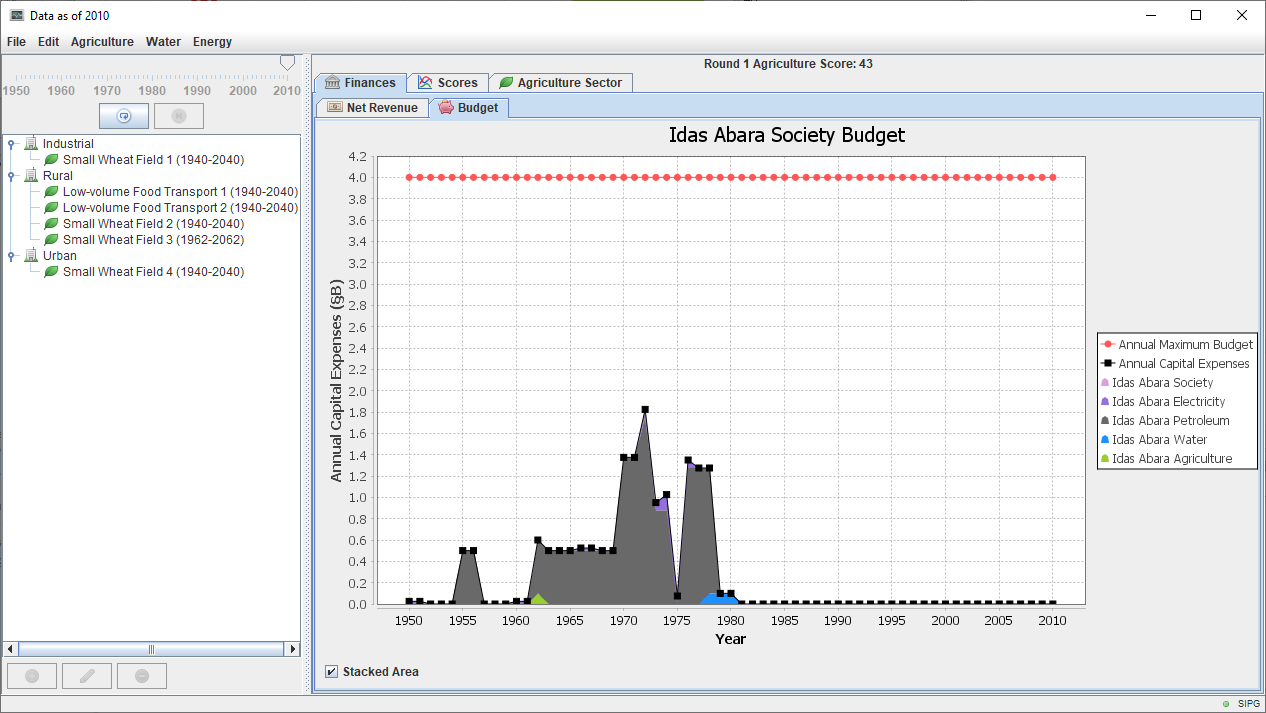}
		\label{fig:gui_budget}}
		
	\subfloat[Sector Net Revenue]{%
		\includegraphics[width=0.48\linewidth,trim={236px 17px 5px 53px},clip]{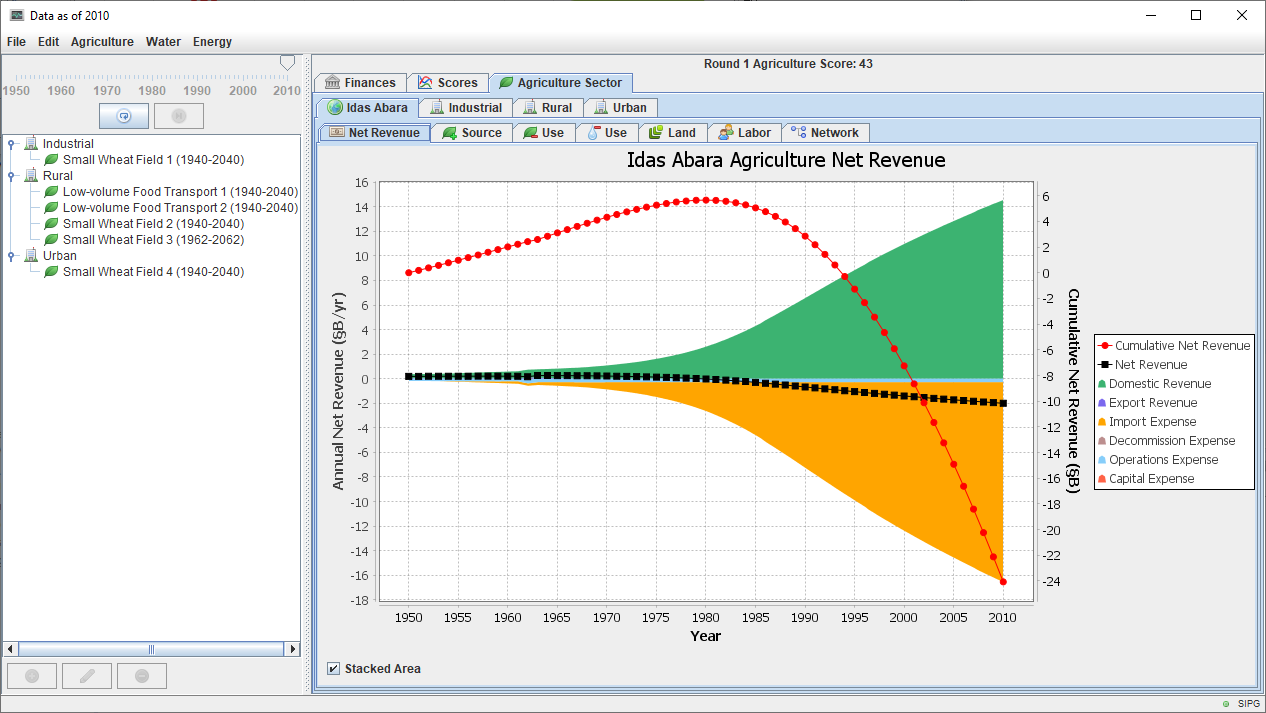}
		\label{fig:gui_sector_revenue}}
	\hfill
	\subfloat[Infrastructure Network Display]{%
		\includegraphics[width=0.48\linewidth,trim={236px 17px 5px 53px},clip]{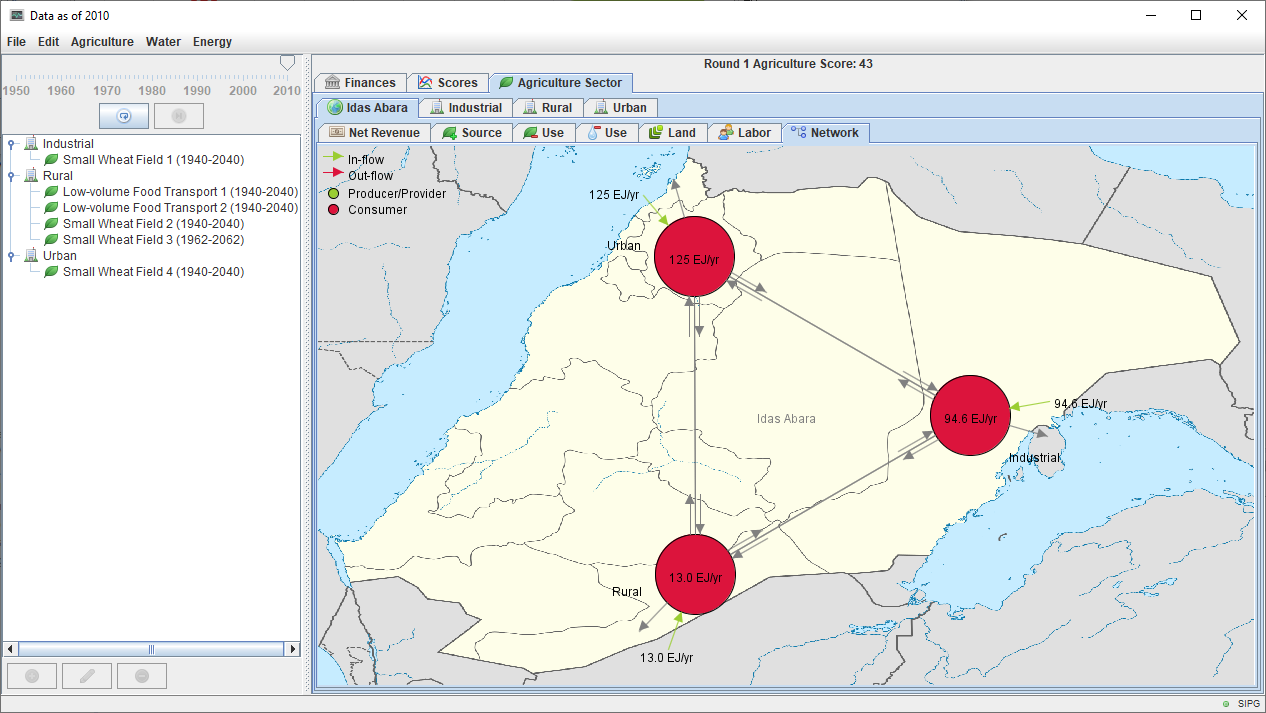}
		\label{fig:gui_network}}
    \caption{Output GUI panels presents societal and sector-specific information for each player role.}
    \label{fig:panel_gui}
\end{figure*}

Simulation outputs are formatted in numerous plots and visualizations in Fig.~\ref{fig:panel_gui} which can be aggregated at the national level or separated by region. Common societal information such as sector-specific contributions to net revenue (Fig.~\ref{fig:gui_net_revenue}) and capital expenditures compared to the annual budget limit (Fig.~\ref{fig:gui_budget}) are available to all players. Other sector-specific information such as net revenue breakdown by source (Fig.~\ref{fig:gui_sector_revenue}), resource sources and sinks (Fig.~\ref{fig:gui_network}), state of natural stocks such as aquifers and reservoirs, and quantified objectives are only visible to individual players.

\subsection{Discussion and Key Limitations}

The SIPG co-simulation artifact technically integrates constituent simulation models while providing decentralized control of constituent models and execution. The co-simulation interface prescribes the types of dynamic dependencies (directed resource exchanges in this scenario) but neither discloses the implementation details of any constituent model nor requires constituent models to be centrally located or shared for execution. These features help overcome legal, proprietary, or simply organizational hurdles to co-design by allowing each design actor to manage and control a component of the technical simulation.

The SIPG scenario and underlying model have been purposefully simplified to facilitate short-duration interactive co-design sessions with non-domain experts. Although many of the underlying concepts are general and could accommodate more realistic or higher-fidelity models, the several key assumptions in Appendix A.5 limit the direct application of results to real-world planning.

Adopting the HLA co-simulation standard carries some technical limitations. First, all constituent members must use the same RTI implementation and the most capable ones are commercially licensed. Further, while there is no strict constraint on model implementation, most RTIs only provide Java and C/C++ language bindings. Additionally, all constituent members must be connected to a common local- or wide-area network configured to allow RTI messages. Finally, any changes to the co-simulation interface, such as adding a new resource dependency, must be documented in the FOM and shared with all members.

The HLA is not the only viable IS architecture for co-design. A broader set of more modern service-oriented and event-driven IS architectures can capably support decentralized information exchange between constituent applications. However, simulation-specific features like time management prove challenging for general-purpose IS software which typically operate under real-time assumptions. Technical considerations of alternative IS architectures is outside the scope of this initial effort to evaluate co-design with co-simulation.

\section{Observational Co-design Study} \label{sec:study}

This section formulates an observational study to evaluate the utility of dynamic exchange of technical information provided by the SIPG co-simulation artifact in a co-design setting. Refined study objectives investigate what game dynamics emerge from interactions between design actors and how co-simulation features influence co-design processes and outcomes.

\subsection{Study Objectives}

As reviewed in Sec.~2, prior literature identifies methods and processes to improve outcomes of engineering design through technical integration and social collaboration. The SIPG scenario exhibits two perspectives on desired outcomes. Each role objective is part of a multi-attribute function where non-dominated solutions form a Pareto-efficient frontier. However, perceived inequity and importance limits desirable solutions to a subset of the efficient frontier that emphasizes synergies across roles. The joint objective identifies synergistic components of individual role objectives that contribute to shared sustainability goals. While it can be conceived of as a holistic single-attribute objective function, in practice, the joint objective only provides effective weighting for role-specific objectives. Thus, from an observer's perspective, desired SIPG outcomes maximize the joint objective but, from a participant's perspective, desired outcomes maximize role-specific and joint objectives as a multi-attribute function with variable weightings from person to person.

To evaluate the SIPG co-simulation artifact's ability to communicate dynamic dependencies using a decentralized IS architecture, this observational study pursues two refined research questions:
\begin{itemize}[label={}]
    \item What game dynamics related to tensions between role-specific and joint objectives in the SIPG scenario emerge from player interactions?
    \item How does dynamic information exchange provided by co-simulation influence technical exchange and pursuit of role-specific and joint objectives?
\end{itemize}

The first question characterizes the dynamical relationships between player roles to identify key information dependencies within the limiting SIPG co-design context. Key resource exchanges such as water for irrigation, electricity for desalination, and currency for capital expenditures are hypothesized to generate conflict between player roles which must be addressed in co-design activities. Analysis of game dynamics compares intermediate and final role-specific and joint objectives to identify tensions in co-design sessions.

Reflecting on SIPG game dynamics that drive a need for co-design, the second question investigates how co-simulation technology influences design activities through dynamic information exchange. Co-simulation is hypothesized to support collective sense-making by communicating resource dependencies as the underlying source of role conflict, allowing role players to identify and manage technical problems within a negotiated solution processes. Analysis compares technical data exchange during design and resulting role-specific and joint outcomes across co-design sessions including variants structured to differentiate co-simulation features.

\subsection{Study Conditions}

This study uses SIPG as the IS platform for a co-design activity. It is structured as a between-subjects study with a design session as the unit of analysis. Two major variants represent co-design environments with strong and weak adoption of co-simulation to probe the effectiveness of dynamic technical interactions. While representing distinct conditions, the study design is not a highly-controlled experiment to understand specific effects of experimental variables on outcomes. Rather, it establishes an observational study to gather initial insights on use of co-simulation technology within a co-design process. The overall study includes 15 co-design sessions equally distributed across three total variants described below.

\begin{figure*}
    \centering
    \includegraphics[scale=1]{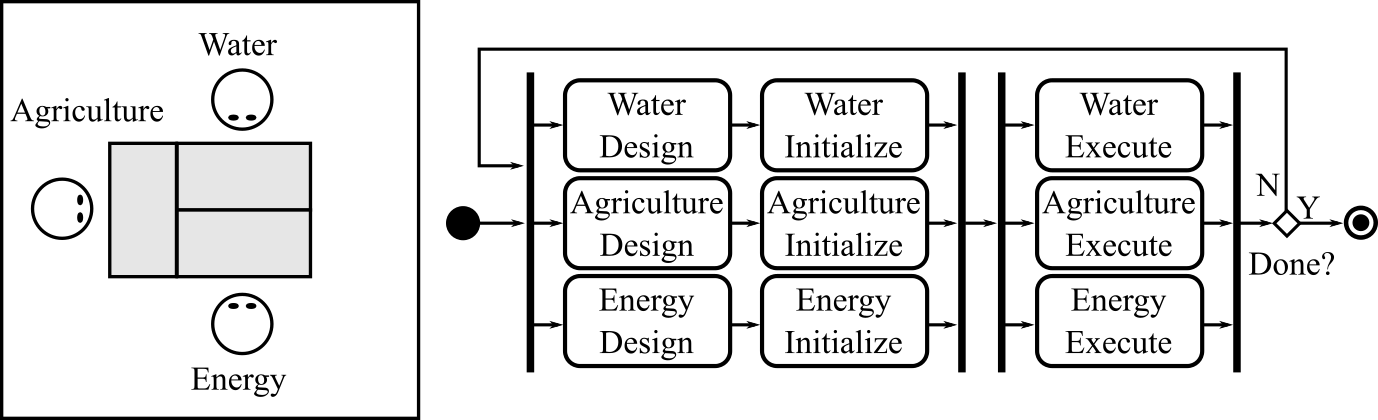}
    \caption{Design station layout and operational activity diagram for Variant 1 which models strong adoption of co-simulation to support dynamic interaction with synchronous exchange and co-location.}
    \label{fig:conditions-1}
\end{figure*}

Variant 1 in Fig.~\ref{fig:conditions-1} models co-design with strong adoption of co-simulation to support dynamic technical interaction. It co-locates the three design stations at a central table and adopts a synchronous mode of information exchange where each participant controls local design inputs but all three must simultaneously run a simulation to update outputs. To trigger a simulation execution, a participant must first click the ``Initialize'' button (loop icon at top left in Fig.~\ref{fig:gui_input_panel}). After all three participants complete initialization, the ``Execute'' button (play icon at top right in Fig.~\ref{fig:gui_input_panel}) unlocks. After all three participants click the ``Execute'' button, the simulation runs on all three design stations in approximately 10--20 seconds, populating analysis results and outputs in Fig.~\ref{fig:panel_gui}.

While not essential for results reported here, Variant 1 sessions disseminate joint objectives either in qualitative or quantitative form. The qualitative form (Variant 1B) identifies the four components of joint objectives (i.e.~food, water, oil, and financial security) in briefing materials. The quantitative form (Variant 1A) provides the joint objective in numeric form that is updated after each execution. Both variants receive role-specific objectives in a quantitative form.

\begin{figure*}
    \centering
    \includegraphics[scale=1]{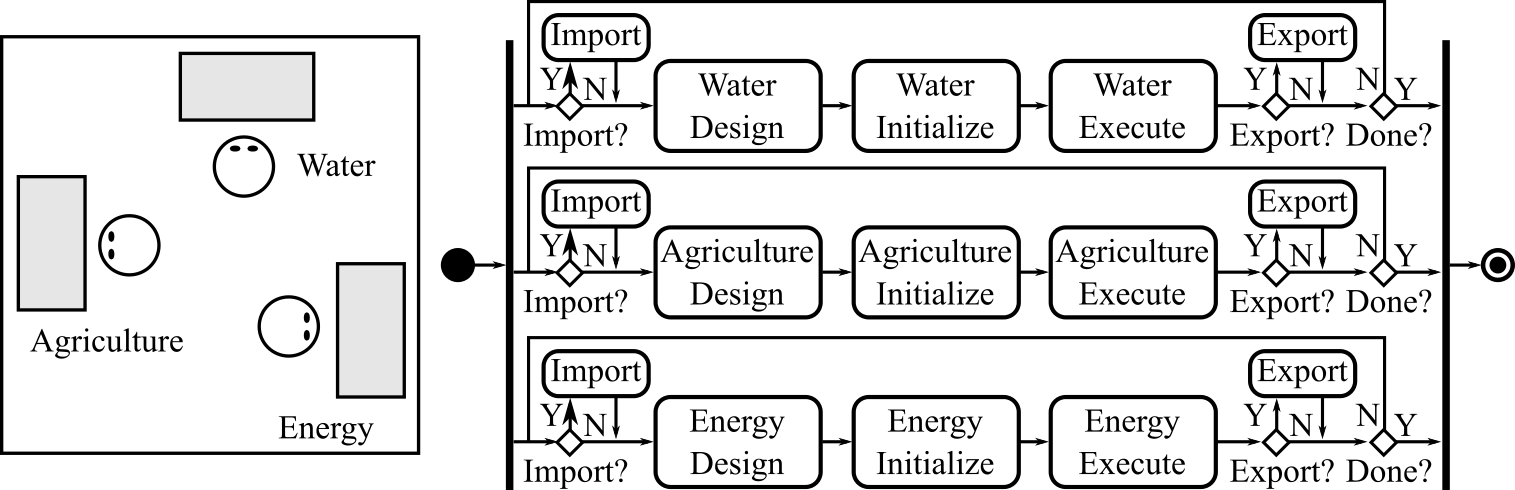}
    \caption{Design station layout and operational activity diagram for Variant 2 which models weak adoption of co-simulation with asynchronous exchange and isolated design stations as barriers to dynamic interaction.}
    \label{fig:conditions-2}
\end{figure*}

Variant 2 in Fig.~\ref{fig:conditions-2} models co-design with weak adoption of co-simulation with mild barriers to dynamic interaction. It isolates design stations at tables several feet apart and adopts an asynchronous mode of information exchange based on importing and exporting static data files. Rather than performing dynamic data exchanges, resource flows at each time step can be saved in a data file and manually transferred between design stations using a shared network folder. Participants retain local control over imports, execution, and exports but moderate discrepancies may arise from out-of-date information. To trigger a new simulation execution after making design changes and/or importing external data, a participant clicks ``Initialize'' and ``Execute'' to immediately populate analysis results. Outputs can later be exported to file format for sharing.

Study observations record intermediate data during the design activity and outcomes at the end of the design activity. Automatic computer logs record all intermediate design decisions and associated meta-information such as timestamps after each simulation execution. While participants are permitted to converse freely and move about the room, no audio or video is recorded. Post-processing tabulates the time and frequency of data exchange actions as process variables, infrastructure plans as design inputs, and objective metrics as design outputs. One data exchange corresponds to a joint simulation execution for co-simulation and new data file availability from all three roles for the asynchronous condition.

\subsection{Study Procedure}

Following an approved protocol (MIT \#1302005518), the study recruited 15 groups of 3 participants from a convenience sample of graduate students without compensation. Table \ref{tab:demographics} summarizes subject demographics. Participants were majority male (64.4\%) and 25--29 years of age (71.1\%) with more college education than work experience. Most participants had never interacted with each other in the past (58.9\% of pairs), although a subset (25.6\%) interact on at least a weekly basis. Inspection yields no significant demographic differences between conditions but the observed demographics may limit generalization of results beyond the sampling frame.

\begin{table}
    \centering
    \caption{Summary of participant demographics.}
    \small
    \begin{tabular}{ll|rrr|rr}
    \hline
    &&\multicolumn{3}{c|}{Variant}&&\\
    Category & Value & 1A & 1B & 2 & Total & (\%) \\
    \hline
    Gender & Male & 11 & 8 & 10 & 29 & 64.4\\
    & Female & 4 & 7 & 5 & 16 & 35.6\\
    \hline
    Age (years) & 18--24 & 3 & 2 & 1 & 6 & 13.3 \\
    & 25--29 & 9 & 13 & 10 & 32 & 71.1 \\
    & 30--34 & 3 & 0 & 4 & 7 & 15.6 \\
    \hline
    College education (years)  & 3--4 & 2 & 1 & 0 & 3 & 6.7 \\
     & 5--6 & 1 & 3 & 4 & 8 & 17.8 \\
    & 7--8 & 7 & 4 & 5 & 16 & 35.6 \\
    & 9+ & 5 & 7 & 6 & 18 & 40.0 \\
    \hline
    Work experience (years) & 0 & 3 & 3 & 6 & 12 & 26.7 \\
    & 1--2 & 9 & 9 & 5 & 23 & 51.1 \\
    & 3--4 & 1 & 3 & 2 & 6 & 13.3 \\
    & 5--6 & 1 & 0 & 2 & 3 & 6.7 \\
    & 7--8 & 1 & 0 & 0 & 1 & 2.2 \\
    \hline
    Interaction with other participants & Never & 19 & 16 & 18 & 53 & 58.9 \\
    & Rarely & 3 & 6 & 3 & 12 & 13.3 \\
    & Monthly & 1 & 1 & 0 & 2 & 2.2 \\
    & Weekly & 5 & 5 & 8 & 18 & 20.0 \\
    & Daily & 2 & 2 & 1 & 5 & 5.6 \\
    \hline
    \end{tabular}
    \label{tab:demographics}
\end{table}

Design sessions are scheduled when three volunteers are available to form ad-hoc groups and are conducted in classrooms. Conditions are assigned in partially-randomized order with the first eight randomly assigned Variant 1A or 1B and the last seven randomly assigned Variant 2 or 1B. At the start of the session, subjects receive a role and color assignment (energy: red, agriculture: green, or water: blue) and sit on one side of a rectangular design station with the fourth seat reserved for the researcher. After introduction, subjects either remain at the central design station (Variant 1) or move to adjacent design stations (Variant 2).

Participants may exit the study at any point, however no such events occurred. A 15-minute scripted presentation introduces the design context including the three regions (industrial, rural, urban), infrastructure within each sector, resource interdependencies, operational behaviors in the simulation model, other assumptions for price and cost, budget and time constraints, and a description of joint objectives. Subjects also receive a confidential sheet describing individual objectives and an overview of key issues in their respective role. Participants may share the confidential information or keep it private. Next, a 15-minute tutorial introduces the subjects to the software tool including simulation inputs (existing elements and available templates), execution control buttons (initialize and run), and a walk-through of all output screens.

After completing addressing any related questions, subjects enter a 60-minute timed design period. Subjects are allowed to move about the room, converse freely with each other, and share their display during the design session, but may not change the room layout. Subjects can also ask the researcher for additional information not displayed in the GUI, clarifications on model assumptions, or other questions excluding advice on design decisions. The researcher updates the remaining time at several points during the session. At the end of the timed period, the researcher leads a de-briefing session to explain the study objectives and probe experiences and observations from the design session.

\subsection{Limitations and Threats to Validity}

This study has several limitations which pose threats to the validity of results. First, it does not employ a highly-controlled design to study context differences between variants. Rather, the two major variants (1 and 2) are only intended to characterize low and high adoption of co-simulation and the two minor variants (1A and 1B) change visibility of joint objectives. More importantly, group processes are largely uncontrolled during design sessions. Subjects are not constrained to follow a particular process for design, nor are there limits on discussion or sharing of information. Furthermore, there are no imposed preferences for individual versus shared objectives. This lack of control introduces additional variation that limits the causal strength of conclusions.

This design does not fully leverage randomization of conditions for practical reasons. Groups are formed as participant schedules allow rather than completely randomly. Potential biases are partially mitigated by the non-purposeful assignment of conditions to sessions which are randomly assigned except for Variant 2 which is limited to the second half of sessions. The ordering effect may bias results due to researcher maturation effects and is partially mitigated by adhering to a common scripted introduction and tutorial across all sessions.

Author participation in the design sessions introduces additional potential biases, especially as subjects are sampled from peer groups and the author is the developer of SIPG. Scripted introduction and tutorial materials and passive participation to only respond to direct questions mitigate some concerns, however the possibility of additional biases must be acknowledged.

Several factors limit the generalizability of results beyond the design sessions considered. Previously discussed limitations in the SIPG model and scenario limit direct extensions to real-world cases. Similarly, the sampled population is not representative of infrastructure planners, although backgrounds in technical areas may be similar. There are also potential reactive effects of experimental arrangements. Participants work in ad-hoc teams and are not required to have background experience with infrastructure systems. Design sessions are conducted in general-purpose classrooms using unfamiliar software tools that requires a large portion of the design time to simply comprehend the task. Finally, participants working in a finite, fictional session may not fully consider the implications of decisions having great socio-economic impact in the real world.

\section{Study Results, Analysis, and Discussion} \label{sec:results}

This section reports results of the observational SIPG study to investigate the game dynamics and how co-simulation features support co-design activities. An overview first describes the collected data and overall features and relationships. Analysis of game dynamics characterizes the tensions between SIPG roles. Additional analysis evaluates how co-simulation features influence design processes and outcomes. Finally, discussion explains the analysis results within the context of SIPG and more broadly for co-design for engineering systems.

\subsection{Overview of Results}

Data post-processing aligns all design decisions on a single timeline to normalize asynchronous and synchronous sessions. Table \ref{tab:sipsgResults} summarizes results at the end of the 60-minute design period for all 15 sessions sorted by variant. Six sessions violated annual budget constraints in one or more years. Inspection shows budget violations are small and isolated to a few years which could be alleviated by adjusting planning schedules to yield similar results. Thus, the final role-specific and joint objective metrics observed are characteristic of valid strategic plans.

\begin{table*}
    \centering
    \footnotesize
    \caption{Summary of design conditions, demographic factors, and outcomes by session.}
    \begin{tabular}{ccccccccllll}
        \hline
        $i$ & Var. & $D_1$ & $D_2$ & $D_3$ & $D_4$ & $D_5$ & Num. & \multicolumn{3}{c}{Role-specific Objective (Rank)} & \multicolumn{1}{c}{Joint Obj.} \\
        \cline{9-11}
        &&&&&&& Exchg. & \multicolumn{1}{c}{Agriculture} & \multicolumn{1}{c}{Water} & \multicolumn{1}{c}{Energy} & \multicolumn{1}{c}{(Rank)} \\
        \hline
        1$^*$ & 1A & 1 & 24.3 & 5.7 & 2.7 & 1.4 & 5 & 624.6 (7) & 352.2 (8) & 602.8 (4) & 344.8 (1) \\
        2$^*$ & 1A & 0 & 26.7 & 8.3 & 1.7 & 0.0 & 6 & 657.7 (9) & 355.2 (10) & 786.9 (14) & 497.9 (11) \\
        3 & 1A & 2 & 22.7 & 6.3 & 0.7 & 12.5 & 10 & 401.1 (2) & 384.0 (14) & 578.4 (1) & 445.1 (5) \\
        4 & 1A & 0 & 24.3 & 5.7 & 2.0 & 0.0 & 11 & 736.3 (12) & 342.7 (3) & 724.4 (10) & 517.1 (15) \\
        5$^*$ & 1A & 1 & 25.0 & 9.0 & 1.0 & 0.0 & 9 & 570.7 (5) & 359.5 (11) & 792.3 (15) & 486.9 (10) \\
        6 & 1B & 1 & 25.0 & 8.3 & 1.7 & 0.0 & 12 & 662.8 (10) & 344.2 (5) & 779.5 (12) & 509.9 (13) \\
        7$^*$ & 1B & 1 & 22.7 & 7.0 & 1.0 & 1.4 & 13 & 613.4 (6) & 312.1 (1) & 688.9 (7) & 484.4 (9) \\
        8$^*$ & 1B & 2 & 22.7 & 6.3 & 0.7 & 0.0 & 5 & 654.1 (8) & 347.8 (7) & 722.4 (9) & 466.9 (7) \\
        9$^*$ & 1B & 2 & 25.0 & 6.3 & 2.0 & 11.1 & 7 & 364.3 (1) & 400.0 (15) & 584.4 (2) & 438.5 (2) \\
        10 & 1B & 1 & 25.0 & 8.3 & 0.7 & 1.6 & 9 & 794.7 (13) & 353.0 (9) & 663.7 (6) & 505.8 (12) \\
        11 & 2 & 0 & 26.7 & 7.0 & 2.3 & 1.4 & 7 & 711.8 (11) & 344.0 (4) & 778.6 (13) & 514.2 (14) \\
        12 & 2 & 2 & 26.7 & 7.7 & 1.7 & 0.1 & 3 & 950.4 (15) & 340.4 (2) & 600.2 (3) & 349.7 (3) \\
        13 & 2 & 2 & 24.3 & 7.0 & 2.0 & 0.0 & 5 & 469.5 (3) & 378.4 (13) & 710.0 (8) & 449.3 (6) \\
        14 & 2 & 0 & 26.7 & 8.3 & 0.7 & 3.7 & 4 & 489.6 (4) & 367.4 (12) & 655.0 (5) & 467.3 (8) \\
        15 & 2 & 1 & 25.0 & 6.3 & 0.3 & 5.8 & 7 & 936.0 (14) & 345.2 (6) & 742.7 (11) & 349.9 (4) \\
        \hline
        \multicolumn{12}{l}{$^*$: Final plan exceeds budget limit in one or more years} \\
        \multicolumn{12}{l}{Demographics: $D_1$: num.~female, $D_2$: avg.~age, $D_3$ avg.~education, $D_4$ avg.~work} \\
        \multicolumn{12}{l}{\hspace{0.9in} $D_5$: avg.~monthly interactions}
    \end{tabular}
    \label{tab:sipsgResults}
\end{table*}

Preliminary analysis performs two-way ANOVA with Python library statsmodels (v0.12.0) function \texttt{anova\_lm} for each aggregate team demographic factor to check for significant effects on outcome observations. Multi-factor effects are not considered due to multicolinearity effects and limited sample size and a Jarque-Bera diagnostic test checks normality assumptions. Age ($D_2$), education ($D_3$), and work experience ($D_4$) aggregated to average team values using lower bounds of item ranges have no significant effect on number of exchanges, role-specific objectives, or joint objectives. Gender as the number of females ($D_1$) shows a significant effect only on the energy objective ($F(1,13)=4.77, p=0.048$) and team familiarity as the average number of interactions with other participants per month ($D_5$) shows a significant effect on the water ($F(1,13)=8.78, p=0.011$) and energy ($F(1,13)=7.06, p=0.020$) role objectives. Additional investigation is required to determine whether gender and team familiarity demographic differences are practically significant.

\subsection{Analysis of Game Dynamics}

As recorded observations are limited to computer logs from simulation executions, this section includes some qualitative observation remarks from the author to supplement quantitative analysis of recorded data.

\begin{figure*}
    \centering
    \includegraphics[scale=1]{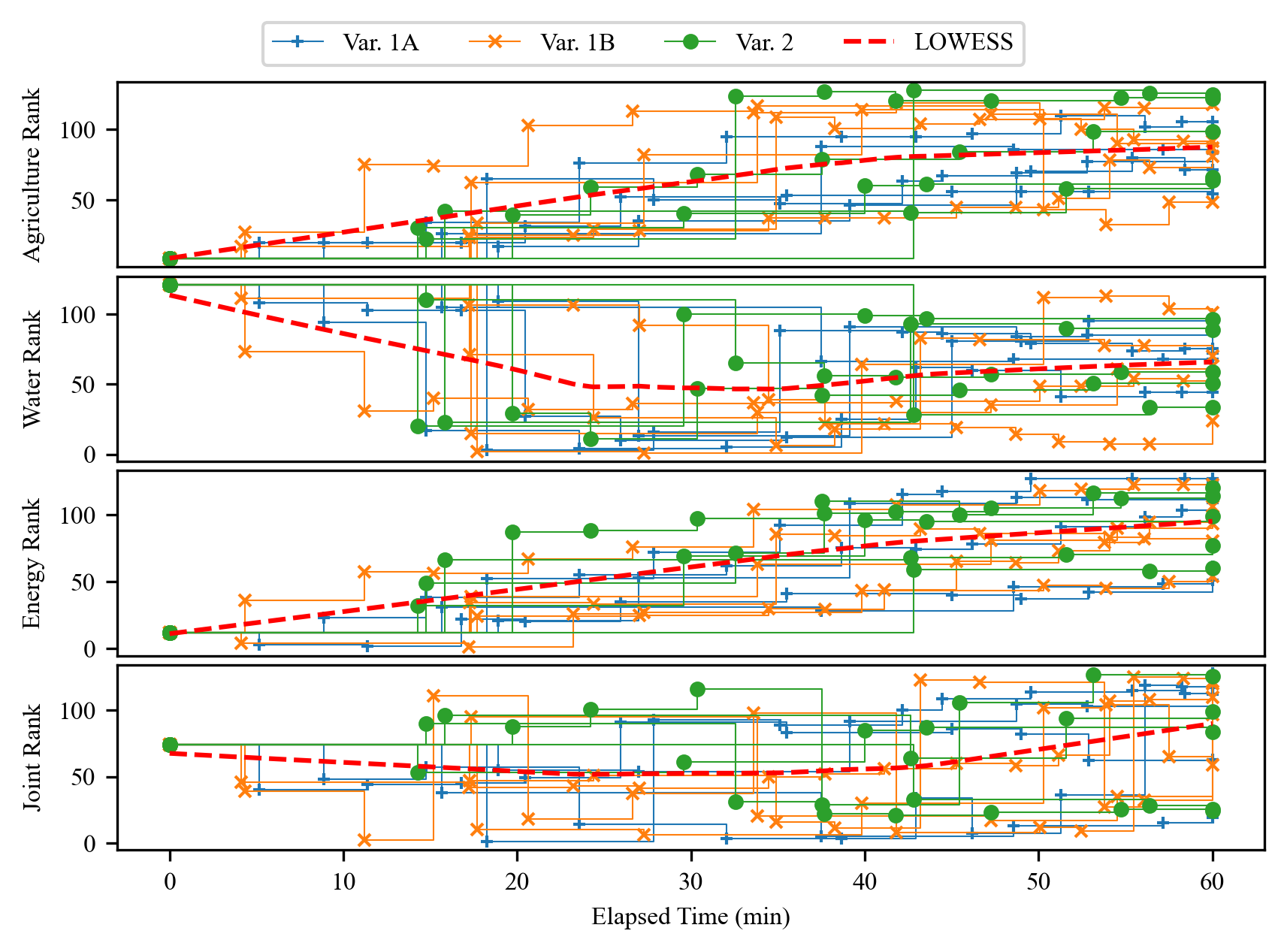}
    \caption{Time series of objective ranks by session variant with locally weighted scatterplot smoothing (LOWESS).}
    \label{fig:history}
\end{figure*}

A time series of role-specific and joint objective ranks after each exchange in Fig.~\ref{fig:history} with a locally weighted scatterplot smoothing (LOWESS) overlay visualizes temporal dynamics. Teams typically spend the opening minutes independently investigating the assigned individual and joint objectives, results of the baseline scenario, and the space of alternative designs. The first exchange often does not occur until 10 or 15 minutes into the 60-minute session. Typical first design changes seek to establish causal understanding within each role. For example, the agriculture player adds food production elements to gauge the magnitude of effect on improving food security or irrigation water demands.

During the first 30 minutes, agriculture and energy role objectives rise but water and joint objectives fall. Players typically follow three strategies during this period: 1) increase food production to improve domestic food security, 2) increase water desalination to improve aquifer security, and 3) increase electricity generation to improve oil consumption efficiency and restore export revenue. Expanding food production improves agriculture and joint objectives but also leads to excess water demands that can reach several multiples of the baseline societal demand. Expanding water production with desalination plants can only satisfy a fraction of the irrigation demand, eventually leading to dire aquifer security consequences for water and joint objectives. Expanding electricity generation with efficient thermal and solar power plants boosts revenue by restoring oil export capacity, increasing the energy and joint objectives.

Activities during the last 30 minutes shift to mitigating huge financial burdens of importing water as the most severe consequences of aquifer depletion. The water player controls information about aquifer health which must be communicated to other players to diagnose and correct structural problems. Additionally, merging plans for costly water and energy infrastructure often require re-phasing to adjust the starting time of planned projects to fit within the annual capital budget limit. Other issues to be addressed during this period include: balancing food production constraints from a limited labor force, expanding food transport capacity between supply and demand regions, balancing revenue from oil export with reservoir security, and expanding oil pipeline capacity between regions. Not all teams have sufficient time to address all issues, contributing to a wide distribution of role-specific and joint objectives.

\begin{figure*}
    \centering
    \includegraphics[scale=1]{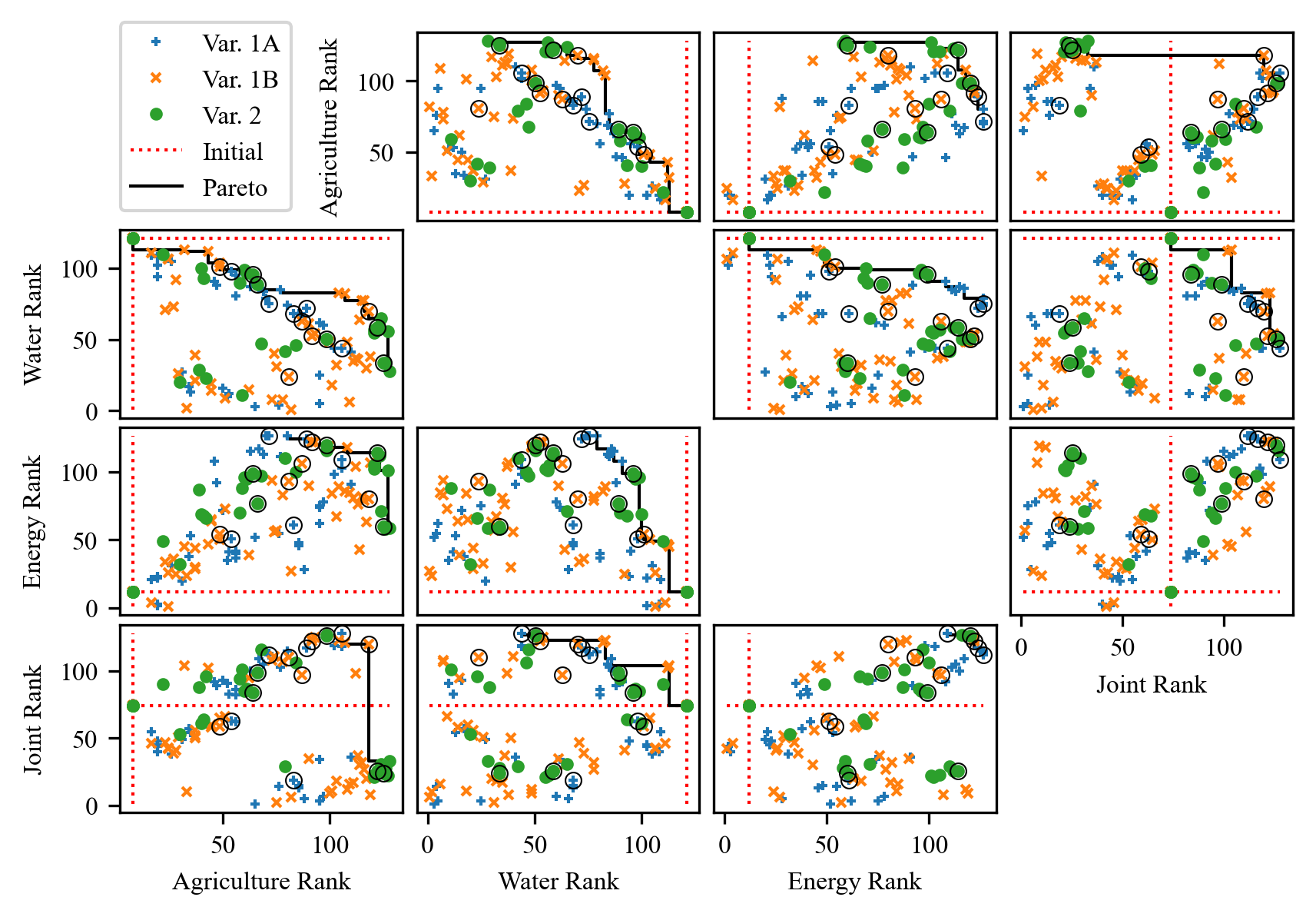}
    \caption{Scatter plot matrix of objective ranks for all design iterations. Black circles annotate final selections.}
    \label{fig:panel}
\end{figure*}

To quantitatively visualize role relationships, Fig.~\ref{fig:panel} shows a scatter plot matrix of role-specific and joint objective ranks for all 128 observed initial, intermediate, and final designs. Final designs close to the Pareto frontier for each pair highlight active constraints as evidence of tension. For example, most final designs fall near the agriculture/water Pareto frontier, indicating a fundamental tension between the two roles linked to irrigation based on the narrative above. Inspection of the role-specific/joint Pareto frontiers shows substantial variability in active constraints across sessions and many dominated points, suggesting the allocated design time was insufficient to converge to a final plan.

Spearman rank sum correlation coefficients computed using Python library SciPy (v1.5.0) function \texttt{spearmanr} in Table~\ref{tab:correlation} summarize role relationships across all 113 observed intermediate and final designs. Agriculture and water role objectives show negative correlations ($p=2.7\cdot10^{-3}$) attributed to irrigation demands on aquifers. Agriculture and energy roles show positive correlations ($p=4.1\cdot10^{-12}$) which is likely due to a confounding factor (i.e., both generally increase over time). Joint objectives show positive correlation with water ($p=0.030$) and energy ($p=2.3\cdot10^{-5}$) role objectives.

\begin{table}
    \centering
    \caption{Spearman correlation coefficients for role-specific and joint objectives.}
    \begin{tabular}{crrr}
        \hline
        & Water & Energy & Joint \\
        \hline
        Agriculture & $-0.280^\dagger$ & $0.594^\ddagger$ & $-0.158$\\
        Water & & $-0.048$ & $0.204^*$ \\
        Energy & & & $0.386^\ddagger$ \\
        \hline
        \multicolumn{4}{c}{$^*: p<0.05, \quad ^\dagger: p<0.01, \quad ^\ddagger: p<0.001$}
    \end{tabular}
    \label{tab:correlation}
\end{table}

\subsection{Analysis of Co-simulation Features}

This section evaluates the effect of co-simulation features on outcomes by comparing major session variants with strong (1) and weak (2) adoption of co-simulation and correlating process and outcome variables across all sessions. Process variables include the number of data exchanges and outcome variables include role-specific and joint objectives.

Initial analysis inspects for differences in process or outcome factors between minor variants (1A and 1B) using a small-sample two-tailed Mann-Whitney U test computed by hand. Results show the minor variant has no statistical effect on data exchanges, role-specific objectives, or joint objectives. Subsequently, observations from Variants 1A and 1B are pooled for further analysis.

Subsequent analysis inspects for differences in process or outcome factors between major variants (1 and 2) using a small-sample two-tailed Mann-Whitney U test computed by hand. Results show only a significant difference in number of data exchanges ($U=8, p=0.04$) between the two variants.

\begin{figure*}
    \centering
    \includegraphics{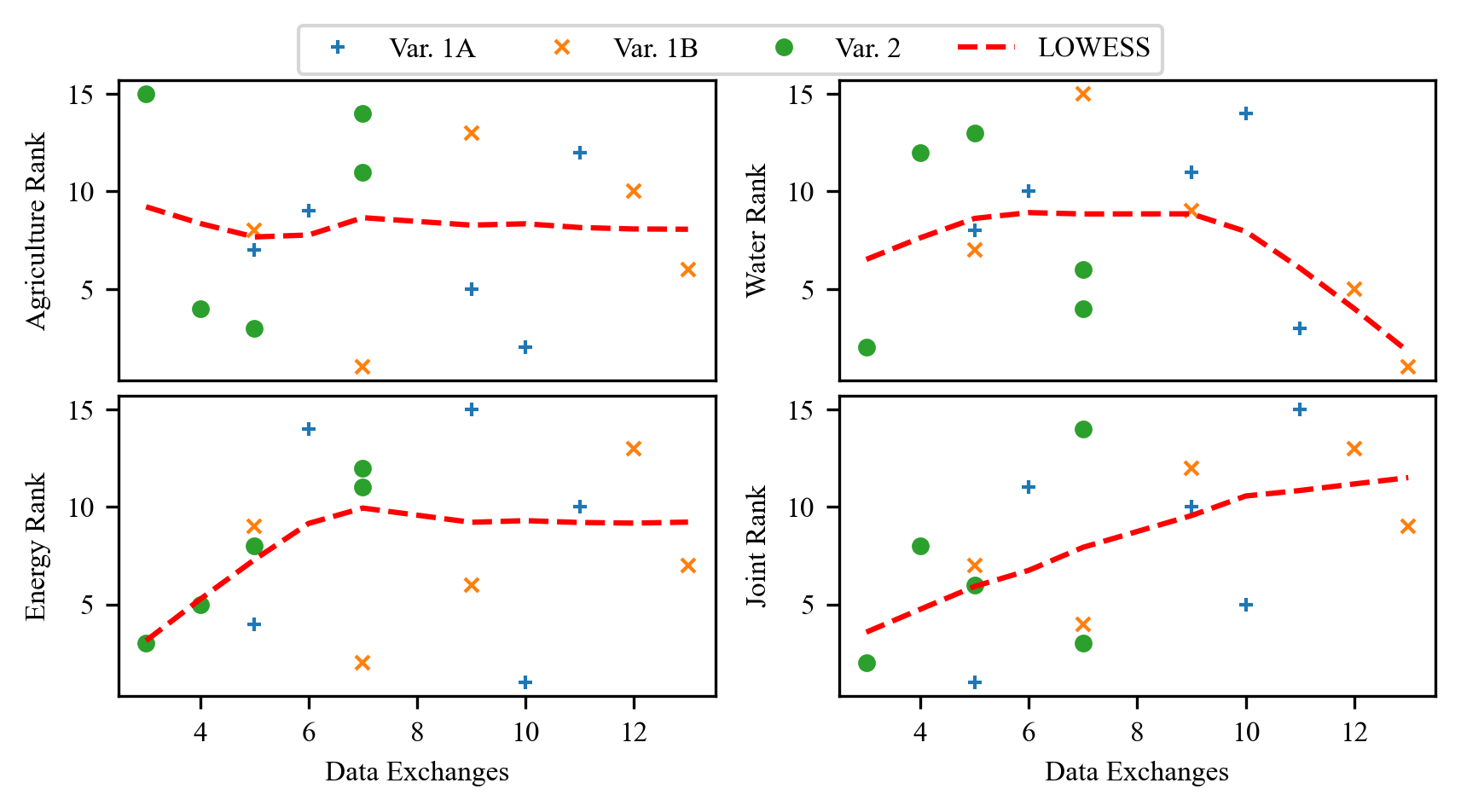}
    \caption{Scatter plot of data exchanges and objective ranks with locally weighted scatterplot smoothing (LOWESS).}
    \label{fig:scatter}
\end{figure*}

Figure~\ref{fig:scatter} compares the number of data exchanges with final objective rank across all sessions with a LOWESS overlay to visualize trends. The joint objective shows the strongest correlation with number of data exchanges, verified by a significant Spearman rank sum correlation coefficient ($r=0.534, p=0.036$) computed using Python library SciPy (v1.5.0) function \texttt{spearmanr}.

\subsection{Discussion of Study Results}

The first study question asked:
\begin{itemize}[label={}]
    \item What game dynamics related to tensions between role-specific and joint objectives in the SIPG scenario emerge from player interactions?
\end{itemize}
Observations and analysis results show strongest tensions between agriculture and water roles driven by water demands for irrigation. A distinct temporal feature shows design actions to increase agricultural production early in a session negatively affect water and joint objectives. Later actions identify and mitigate the water resource problem to partly recover both water and joint objectives. This game dynamic is particularly challenging because the water role player controls the critical information about aquifer state but has limited ability to independently address the problem by increasing desalination capacity. In contrast, the energy role has no similarly strong tensions. Design actions to expand energy infrastructure directly contribute to both role-specific and joint objectives. While desalination capacity contributes to electricity demand, there are no similar dependencies on energy resource flows for other roles.

The observed SIPG game dynamics share some similarities with the historical case of Saudi Arabia \citep{DeNicola_2015_ClimateChangeWaterScarcity}. Agricultural expansion in the 1980s significantly increased groundwater withdrawals from aquifers. Desalination projects, while largest in the world, only contribute a small fraction of total water demands. Policy efforts over the past decade have moved to reduce agricultural production and import water-intense products as a type of ``virtual water'' but there remain significant challenges to sustain the rapidly growing and urbanizing population. Complicating the historical comparison, scientific knowledge about aquifers has improved dramatically since the 1980s, contributing to changing understanding of sustainability implications of their depletion. Nevertheless, parallels to water as a focal role with limited independent ability to change outcomes helps establish importance of dynamic information exchanges.  

The second study question asked:
\begin{itemize}[label={}]
    \item How does the dynamic information exchange provided by co-simulation influence technical exchange and role-specific and joint objectives?
\end{itemize}
Two major variants (1 and 2) modify co-design settings to represent strong and weak adoption of co-simulation. While both variants exchange dynamic resource dependencies using simulation, Variant 1 is synchronous and Variant 2 is asynchronous. From a practical perspective, Variant 1 simplifies data exchange by running a single co-simulation execution but also couples design iterations across roles by requiring group consensus to run. In contrast, Variant 2 simplifies simulation execution by decoupling role-specific design iterations but requires more steps to export and import static dependency files. Analysis results show Variant 1 sessions have a larger number of data exchanges than Variant 2; however, it cannot be determined whether this difference is driven by ease of use or simply because co-simulation is required to observe simulation results.

Looking across all design sessions, analysis results show technical data exchange is positively correlated with joint objective outcomes. While causation cannot be determined, it is plausible that data exchange helps to compute and identify the resource dependencies underlying tensions between roles. More frequent data exchange (up to a limit) may help identify sources of poor-performing designs and convene negotiation activities to propose alternatives. Synchronous co-simulation in Variant 1 facilitates technical data exchange but cannot statistically be linked to better joint objective outcomes, in part due to small sample sizes and high outcome variability.

\begin{figure*}
    \centering
    \includegraphics{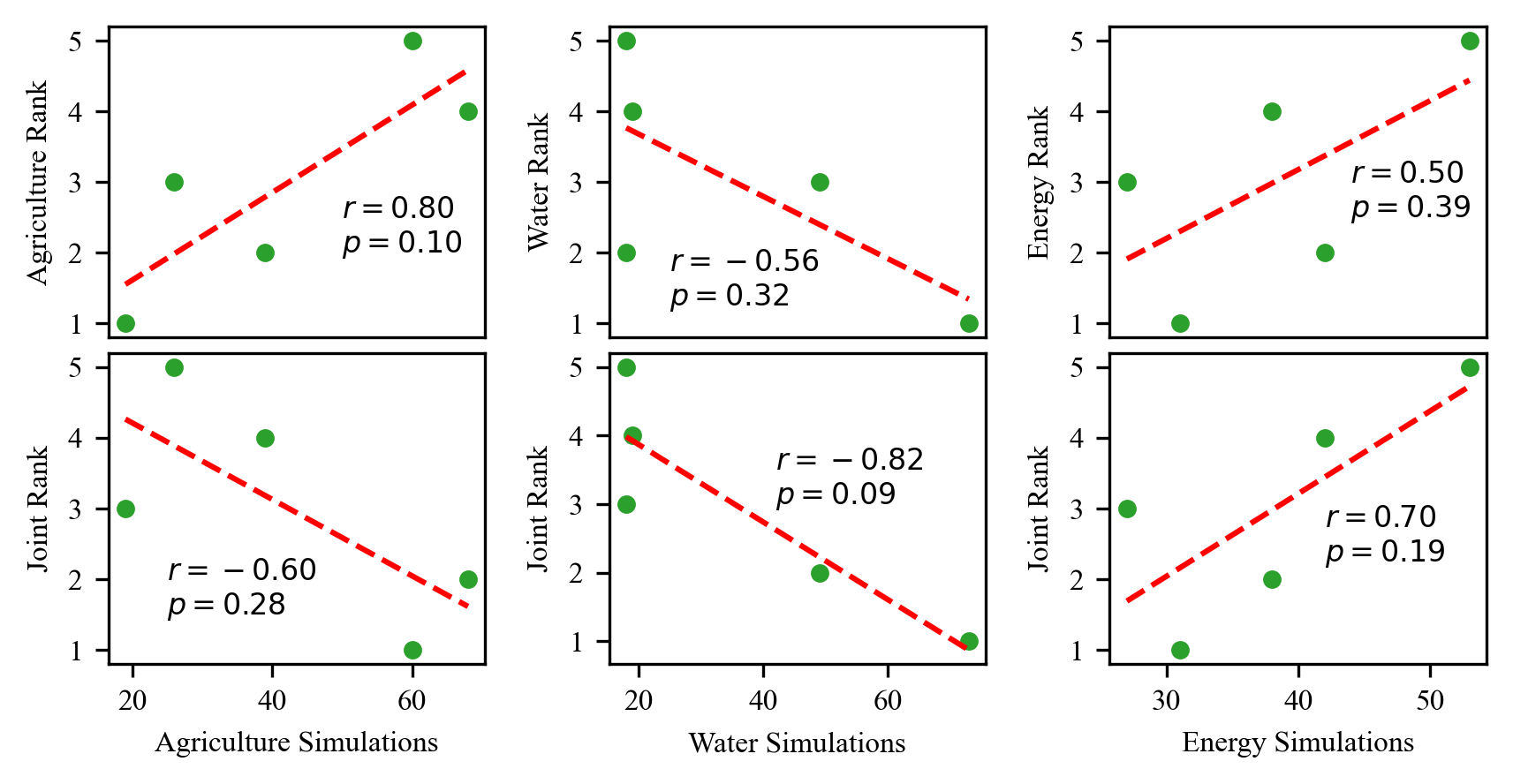}
    \caption{Scatter plot of asynchronous simulations conducted and objective ranks for Variant 2 sessions with linear trend overlay and Spearman rank sum correlation annotation (with $p$-value).}
    \label{fig:scatter-2}
\end{figure*}

Table~\ref{tab:simex} inspects the number of simulations conducted for Variant 2 sessions to further investigate design activities without dynamic exchange from co-simulation. Figure~\ref{fig:scatter-2} illustrates linear trends between simulations conducted and resulting objective ranks (note that rank sum correlations are large but not statistically significant due to small sample size). The ability to conduct independent simulations for roles like the agriculture player may indeed harm pursuit of joint objectives because they obscure key insights (e.g., aquifer depletion) by referencing static dependencies and further delay its acquisition by expending effort on local objective maximization rather than coordination and data exchange. In contrast, decoupled simulation may help more independent roles like the energy player achieve individual and joint objectives.

\begin{table*}
    \small
    \centering
    \caption{Summary of role-specific simulation executions and objective ranks in asynchronous sessions.}
    \begin{tabular}{cccccllll}
        \hline
        $i$ & \multicolumn{1}{c}{Num.} & \multicolumn{3}{c}{Simulations Conducted} & \multicolumn{3}{c}{Role Objective (Rank)} & \multicolumn{1}{c}{Joint Obj.} \\
        \cline{3-5}\cline{6-8}
        &  \multicolumn{1}{c}{Exchg.} & Agriculture & Water & Energy & Agriculture & Water & Energy & \multicolumn{1}{c}{(Rank)} \\
        \hline
        11 & 7 & 26 & 18 & 53 & 711.8 (3) & 344.0 (2) & 778.6 (5) & 514.2 (5) \\
        12 & 3 & 60 & 73 & 31 & 950.4 (5) & 340.4 (1) & 600.2 (1) & 349.7 (1) \\
        13 & 5 & 19 & 18 & 27 & 469.5 (1) & 378.4 (5) & 710.0 (3) & 449.3 (3) \\
        14 & 4 & 39 & 19 & 42 & 489.6 (2) & 367.4 (4) & 655.0 (2) & 467.3 (4) \\
        15 & 7 & 68 & 49 & 38 & 936.0 (4) & 345.2 (3) & 742.7 (4) & 349.9 (2) \\
        \hline
    \end{tabular}
    \label{tab:simex}
\end{table*}

Finally, this discussion must mention the many limitations of above results. Similar to other research applications of simulation gaming, establishing highly-controlled and large-sample human studies with rich contextual applications is challenging and often impossible. The discussion provided here balances quantitative conclusions (where possible) with observational insights to inform future work. Some of the most important limitations relate to unobserved or uncontrolled factors that contribute to large observed variation. For example, the recorded computer logs only contain a fraction of the communication between role players and more complete records would capture conversations and movements about the room to observe collaborative activities. Additionally, stronger controls on team demographic factors would help understand differential effects of domain knowledge, team familiarity, communication style, and a myriad of other factors known to influence group work. The relatively high task complexity of the SIPG scenario and limited session duration may also benefit from estimates of cognitive workload to understand group-level differences in comprehension and control exerted over the design problem.

\subsection{Implications for Co-design with Co-simulation}

Beyond the SIPG scenario-specific questions above, the observational study evaluates co-simulation in a co-design setting to answer the top-level question:
\begin{itemize}[label={}]
    \item How can co-simulation artifacts technically integrate and provide dynamic information exchange among design actors during co-design activities?
\end{itemize}
The technical design of the SIPG co-simulation artifact identifies directed resource flows as the primary dependency between design actors. Synchronous co-simulation exchanges dynamic resource dependencies and computes their effects on design outcomes as a precursor to negotiated solution processes. In contrast, co-design without co-simulation can adopt technical solution processes like local objective optimization in the absence of structured collaborative processes. Differences in co-design settings appear most important when roles exhibit strong conflicts or tensions, such as the agriculture and water roles in SIPG. The higher cost of co-simulation may only be beneficial in such settings where it exposes critical dependencies.

Quantitative conclusions of the observational study are limited due to high outcome variability and imperfect comparison cases to truly evaluate the effect of co-simulation. Results emphasize a process-oriented perspective by evaluating differences in the co-design activity like the number of data exchanges as a type of coordination behavior associated with desired joint outcomes. Future studies should seek to provide additional evidence for causal links between process and outcome factors to focus IS development on supporting specific processes.

Finally, results from the SIPG scenario represent only a narrow perspective on co-design focused on engineering decision-making. Results may not generalize to other types of co-design settings with different types of game dynamics and tensions among player roles. Furthermore, other co-design applications such as participatory integrated assessment and simulation gaming frequently consider a broader set of design actors including policy-makers, consumers, and community members who play different roles than decision decision-makers yet also represent key constituents in infrastructure planning.

\section{Conclusion} \label{sec:conclusion}

Co-design spans both technical integration and social negotiation perspectives important to address engineering systems challenges of societal significance. Co-simulation encompasses a type of supporting IS that permits dynamic information exchange between design actors. Co-design benefits from co-simulation as a source of technical information structured within in a social activity that can make visible issues of shared interest in pursuit of joint objectives.

This work demonstrates a co-simulation artifact for a strategic infrastructure planning co-design scenario that draws some parallels to Saudi Arabia between 1950 and 2010. Three player roles control agriculture, water, and energy (electricity and petroleum) infrastructure to satisfy demands of a non-player societal role. The co-simulation artifact exchanges directed resource flows as players design new infrastructure in pursuit of role-specific and joint objectives.

Co-simulation standards such as the HLA provide a technical means to couple simulation executions but also require technical coordination to align temporal and spatial scales across constituent models. The SIPG co-simulation artifact adopts a graph-based framework with nodes to aggregate spatial resources and annual time steps to aggregate temporal behaviors and iteration to resolve interdependencies. While generally applicable to resource infrastructure systems, this work does not fully address challenges of integrating disparate spatial or temporal scales and coping with high cost and complexity of co-simulation.

An observational study using the SIPG artifact conducted 15 one-hour design sessions. Two session variants model strong and weak adoption of co-simulation with synchronous and asynchronous modes of simulation execution, respectively. Observed game dynamics emphasize conflict and tension between agriculture and water players regarding irrigation resource demands while the energy player remains more independent. Analysis across sessions show that co-simulation variants exhibit more numerous data exchanges and, across all sessions, data exchange is positively correlated with higher achievement of joint objectives.

While the results of this study are limited by the prototypical nature of the SIPG artifact and limited sample size with non-domain experts, they provide initial support for co-simulation applied to a co-design setting. Future work should explore alternative IS implementations to support co-simulation with reduced cost and complexity compared to the HLA. Even at relatively low levels of fidelity or detail, co-simulation activities have the potential to anchor co-design sessions with exchange of technical data and contribute to the joint inquiry and imagination necessary to address critical contemporary issues facing society.

\section*{Acknowledgement}

This work was supported in part by a National Defense Science and Engineering Graduate (NDSEG) fellowship with equipment support (computers and HLA RTI license) from the Center for Complex Engineering Systems at the Massachusetts Institute of Technology.

\appendix
\section{Model Implementation Details} 
\label{sec:model_details}

This section provides details on the sector-specific infrastructure system models. The notation $\mathcal{E}(n)$ gives the set of infrastructure elements originating at node $n\in\mathcal{N}$ and $\mathcal{E}^\prime(n)$ gives the set of elements terminating at node $n\in\mathcal{N}$.

\subsection{Societal System Model}

The non-player societal role generates regional demands for food, water, oil, and electricity as a function of population. Table \ref{tab:societal_properties} defines societal model properties for each node, loosely selected fit the historical context of Saudi Arabia (in aggregate) between 1950 and 2010.

\begin{table*}
    \footnotesize
    \caption{Societal system node properties.}
    \begin{tabular}{cclcccl}
        \hline
        Resource & Variable & Description & Industrial & Urban & Rural & Units \\
        \hline
        Population & $t_0^{pop}$ & Datum time & 1980 & 1980 & 1980 & year \\
        & $P_0$ & Datum population & 3.0 & 6.0 & 0.75 & million people \\
        & $P_{max}$ & Maximum population & 17.5 & 20.0 & 4.0 & million people \\
        & $r_{pop}$ & Logistic growth rate & 7 & 6 & 5 & \% \\
        \hline
        Food & $t_0^{food}$ & Datum time & 1975 & 1975 & 1975 & year \\
        & $d_{0}^{food}$ & Datum per-capita demand & 2300 & 2300 & 2300 & kcal/day \\
        & $d_{min}^{food}$ & Minimum per-capita demand & 1800 & 1800 & 1800 & kcal/day \\
        & $d_{max}^{food}$ & Maximum per-capita demand & 5800 & 5800 & 5800 & kcal/day \\
        & $r_{food}$ & Logistic growth rate & 20 & 20 & 20 & \% \\
        \hline
        Water & $t_0^{water}$ & Datum time & 1965 & 1965 & 1965 & year \\
        & $d_{0}^{water}$ & Datum per-capita demand & 175 & 175 & 175 & L/day \\
        & $d_{min}^{water}$ & Minimum per-capita demand & 25 & 25 & 25 & L/day \\
        & $d_{max}^{water}$ & Maximum per-capita demand & 325 & 325 & 325 & L/day \\
        & $r_{water}$ & Logistic growth rate & 8 & 8 & 8 & \% \\
        \hline
        Oil & $t_0^{oil}$ & Datum time & 1970 & 1970 & 1970 & year \\
        & $d_{0}^{oil}$ & Datum per-capita demand & 1 & 1 & 1 & toe/year \\
        & $d_{min}^{oil}$ & Minimum per-capita demand & 0 & 0 & 0 & toe/year \\
        & $d_{max}^{oil}$ & Maximum per-capita demand & 9 & 9 & 9 & toe/year \\
        & $r_{oil}$ & Logistic growth rate & 9 & 9 & 9 & \% \\
        \hline
        Electricity & $t_0^{elect}$ & Datum time & 1950 & 1950 & 1950 & year \\
        & $d_{0}^{elect}$ & Datum per-capita demand & 0.25 & 0.25 & 0.25 & kWh/day \\
        & $d_{min}^{elect}$ & Minimum per-capita demand & 0 & 0 & 0 & kWh/day \\
        & $d_{max}^{elect}$ & Maximum per-capita demand & 40 & 40 & 40 & kWh/day \\
        & $r_{elect}$ & Logistic growth rate & 9 & 9 & 9 & \% \\
        \hline
        \multicolumn{7}{l}{Units: kcal: kilocalorie (Calorie); L: liter; toe: ton of oil equivalent; kWh: kilowatt hour} 
    \end{tabular}
    \label{tab:societal_properties}
\end{table*}

A logistic growth function models population growth in each region, parameterized by a datum population $P_0$ at time $t_0$, a maximum long-term population (carrying capacity) $P_{max}$, and logistic growth rate $r_p$. The population of region $n$ in year $t$ is given by
\begin{equation}
    P(n,t) = \frac{P_{max}(n)\cdot P_0(n) \cdot e^{r_p(n)\cdot(t-t_0(n))}}{P_{max}(n) + P_0(n)\cdot\left(e^{r_p(n)\cdot(t-t_0(n))}-1\right)}.
\end{equation}

A logistic function also models growth in per-capita resource demands, parameterized by a minimum demand $d_{min}$, maximum demand $d_{max}$, datum demand $d_{0}$ at time $t_0$, and logistic growth rate $r$. The per-capita demands for resource of type $\tau$ in region $n$ at time $t$ is given by
\begin{equation}
    d_\tau(n,t) = d_{min}^\tau(n) + \frac{\left(d_{max}^\tau(n)-d_{min}^\tau(n)\right)\cdot\left(d_0^\tau(n)-d_{min}^\tau(n)\right)\cdot e^{r_\tau(n)\cdot(t-t_0^\tau(n))}}{\left(d_{max}^\tau(n)-d_{min}^\tau(n)\right)+\left(d_0^\tau(n)-d_{min}^\tau(n)\right)\cdot\left(e^{r_\tau(n)\cdot(t-t_0^\tau(n))}-1\right)}
\end{equation}
such that the societal demand for resource $\tau$ in region $n$ at time $t$ is $D^{societal}_\tau(n)=P(n,t)\cdot d_\tau(n,t)$.

Additionally, the societal system aggregates net revenues from each of the other system models. Sector-specific revenues include domestic and export sales. Expenses include domestic and import purchases as well as capital and operations costs. The currency stock is updated at each time step using net revenues from each regional infrastructure
\begin{equation}
    Q_{currency}(t+\Delta t) = Q_{currency}(t) + \sum_{n \in \mathcal{N}} \left( Q^{agricul}_{currency}(n) +  Q^{water}_{currency}(n) + Q^{energy}_{currency}(n) \right).
\end{equation}

\subsection{Agriculture System Model}

Agriculture system properties in Table~\ref{tab:ag_properties} define prices for domestic, imported, and exported food resources and set the workforce participation and arable land area for each node. The rural region has the largest workforce fraction and arable land area but its low population does not demand as much food as other regions, presenting a logistical challenge for distribution.

\begin{table*}
    \footnotesize
    \caption{Agriculture system node properties.}
    \begin{tabular}{clcccl}
        \hline
        Variable & Description & Industrial & Urban & Rural & Units \\
        \hline
        $\pi^{local}_{food}$ & Price of domestic food (no net trade impact) & 60 & 60 & 60 & $\S$/GJ \\
        $\pi^{import}_{food}$ & Price of imported food (net trade deficit) & 70 & 70 & 70 & $\S$/GJ \\
        $\pi^{export}_{food}$ & Price of exported food (net trade surplus) & 50 & 50 & 50 & $\S$/GJ \\
        $f^{labor}_{pop}$ & Maximum labor workforce participation & 4 & 4 & 40 & \% \\
        $q_{land}$ & Arable land area & 8 & 10 & 15 & thousand km$^2$\\
        \hline
        \multicolumn{6}{l}{Units: $\S$: fictional currency; GJ: gigajoule; km: kilometer}
    \end{tabular}
    \label{tab:ag_properties}
\end{table*}

Agriculture element properties in Table~\ref{tab:ag_elements} define two sizes of fields to produce food and two sizes of roads to transport food between regions. Larger infrastructure benefit from slight economies of scale. Players instantiate infrastructure from these templates to design a strategic plan.

\begin{table*}
    \footnotesize
    \caption{Agriculture system element properties.}
    \begin{tabular}{clccccl}
        \hline
        Variable & Description & Sm.~Field & Lg.~Field & Sm.~Road & Lg.~Road & Units  \\
        \hline
        $p_{capital}$ & Capital expense & 100 & 180 & 50 & 300 & million $\S$/year \\
        $d_{capital}$ & Capital expense duration & 1 & 1 & 1 & 1 & year \\
        $p_{fixed}$ & Fixed expense & 5 & 9 & 2.5 & 15 & million $\S$/year \\
        $f_{land}^{currency}$ & Variable expense (field) & 50000 & 45000 & -- & -- & $\S$/km$^2$/year \\
        $q_{max}^{land}$ & Maximum land area & 500 & 1000 & -- & -- & km$^2$ \\
        $f_{land}^{labor}$ & Land-labor intensity & 60 & 60 & -- & -- & person/km$^2$ \\
        $f_{land}^{water}$ & Land-water intensity & 1.5 & 1.5 & -- & -- & MCM/km$^2$/year \\
        $f_{land}^{food}$ & Land-food productivity & 5 & 5 & -- & -- & TJ/km$^2$/year \\
        $f_{food}^{currency}$ & Variable expense (road) & -- & -- & 2 & 2 & $\S$/GJ \\
        $q_{max}^{transport}$ & Maximum throughput & -- & -- & 2 & 15 & EJ/year \\
        $\eta$ & Transport efficiency & -- & -- & 92 & 94 & \% \\
        \hline
        \multicolumn{7}{l}{Units: $\S$: fictional currency; MCM: million cubic meters; GJ: gigajoule; TJ: terajoule; EJ: exajoule}
    \end{tabular}
    \label{tab:ag_elements}
\end{table*}

The agriculture system is coupled with societal and water systems. Regional food demand arises from the societal system, $D_{food}(n) = D^{societal}_{food}(n)$. Water resources required to satisfy the operational plan at each node total
\begin{equation}
    D^{agricul}_{water}(n) = \sum_{e \in \mathcal{E}(n)} f_{land}^{water}(e) \cdot q^{use}_{land}(e).
\end{equation}
The agriculture system controller sets food production and transport levels in constituent infrastructure elements and determines the quantity of food to import and export at each node by solving the following linear program:

\noindent Find:
\begin{align}
    & q^{use}_{land}(e), q^{transport}_{food}(e) \; \forall \; e \in \mathcal{E} \\
    & q_{food}^{import}(n), q_{food}^{export}(n) \; \forall \; n \in \mathcal{N}
\end{align}
Minimize:
\begin{align}
    \begin{split}
        & \sum_{e \in \mathcal{E}} \left( f^{currency}_{land}(e)+f^{water}_{land}(e)\cdot \pi_{water}^{local} \right)\cdot q^{use}_{land}(e) + f^{currency}_{transport}(e)\cdot q^{transport}_{food}(e) \\
        &+ \sum_{n\in\mathcal{N}} \left( \pi_{food}^{import} \cdot q_{food}^{import} (n) - \pi_{food}^{export} \cdot q_{food}^{export} (n) \right)
    \end{split} 
\end{align}
Subject to:
\begin{align}
    & q^{use}_{land}(e) \leq q^{land}_{max}(e) \; \forall \; e \in \mathcal{E} \\
    & q^{transport}_{food}(e) \leq q^{transport}_{max}(e) \; \forall \; e \in \mathcal{E} \\
    & \sum_{e \in \mathcal{E}(n)} q^{use}_{land}(e) \leq q_{land}(n) \; \forall \; n \in \mathcal{N} \\
    & \sum_{e \in \mathcal{E}(n)} f^{labor}_{land}(e) \cdot q^{use}_{land}(e) \leq f^{labor}_{pop}(n)\cdot P(n) \; \forall \; n \in \mathcal{N} \\
    \begin{split}
        & \sum_{e \in \mathcal{E}(n)} \left( f_{food}^{land}(e)\cdot q^{use}_{land}(e)-q^{transport}_{food}(e) \right) + \sum_{e \in \mathcal{E}^\prime(n)}  \eta(e)\cdot q^{transport}_{food}(e) \\
        & + q_{food}^{import} (n) - q_{food}^{export}(n) \geq D_{food}(n) \; \forall \; n \in \mathcal{N}
    \end{split}
\end{align}

Net revenues accumulated by regional agriculture systems include revenue from local, regional distribution, and export sales, resource expenses from regional distribution and import purchases, and other expenses from capital, fixed, and/or variable costs based on lifecycle phase
\begin{equation}
    \begin{split}
        Q_{currency}^{agricul}(n) = & \pi^{local}_{food} \cdot D_{food}(n) + \sum_{e \in \mathcal{E}(n)} \left( \pi^{local}_{food} \cdot \eta(e) \cdot q^{transport}_{food}(e) \right) + \pi^{export}_{food} \cdot q^{export}_{food}(n) \\
        & - \sum_{e \in \mathcal{E}^\prime(n)} \left( \pi^{local}_{food} \cdot \eta(e) \cdot q^{transport}_{food}(e) \right) - \pi^{import}_{food} \cdot q^{import}_{food}(n) \\
        & - \sum_{e \in \mathcal{E}(n)}\left( p_{capital}(e) + p_{fixed}(e) + p_{variable}(e) \right)
    \end{split} 
\end{equation}
where variable costs include operating expenses and consumption of water resources
\begin{equation}
    p_{variable}(e) = \left( f_{land}^{currency}(e) + \pi^{local}_{water}\cdot f_{land}^{water}(e) \right) \cdot q^{use}_{land}(e) + f_{food}^{currency}(e) \cdot q^{transport}_{food}(e).
\end{equation}

\subsection{Water System Model}

Water system properties in Table~\ref{tab:wa_properties} define prices for domestic and imported, water resources, determine coastal access for desalination, set the initial stock in 1950 and recharge rate of aquifers, and set resources required to lift water at each node. All three regions have small recharge rates relative to the initial volume, representative of largely non-renewable sources. Although aquifers increase in salinity under heavy withdrawal, water quality is not considered in this model and the aquifers are assumed to produce potable water until completely depleted.

\begin{table*}
    \footnotesize
    \caption{Water system node properties.}
    \begin{tabular}{clcccl}
        \hline
        Variable & Description & Industrial & Urban & Rural & Units \\
        \hline
        $\pi^{local}_{water}$ & Price of domestic water (no net trade impact) & 0.05 & 0.05 & 0.05 & $\S$/m$^3$ \\
        $\pi^{import}_{water}$ & Price of imported water (net trade deficit) & 10 & 10 & 10 & $\S$/m$^3$ \\
        $Q^{aquifer}_{0}$ & Initial aquifer volume & 200 & 150 & 250 & km$^3$ \\
        $r_{recharge}$ & Aquifer recharge rate & 0.1 & 2.2 & 1.2 & km$^3$/year \\
        $b_{coastal}$ & Coastal access for desalination & 1 & 1 & 0 & -- \\
        $f_{water}^{aquifer}$ & Lifting aquifer intensity & 1.0 & 1.0 & 1.0 & m$^3$/m$^3$ \\
        $f_{water}^{elect}$ & Lifting electrical intensity & 0.9 & 0.9 & 0.9 & kWh/m$^3$ \\
        \hline
        \multicolumn{6}{l}{Units: $\S$: fictional currency; km: kilometer; kWh: kilowatt hour}
    \end{tabular}
    \label{tab:wa_properties}
\end{table*}

Water element properties in Table~\ref{tab:wa_elements} define three sizes of desalination plants modeled based on reverse osmosis technology. This process is energy-intensive, requiring more than four times the electricity of comparatively-simple aquifer lifting. Note that even the largest desalination capacity (0.6 km$^3$/year) represents only a small fraction of the aquifer volume.

\begin{table*}
    \footnotesize
    \caption{Water system element properties.}
    \begin{tabular}{clcccl}
        \hline
        Variable & Description & Small Desal. & Large Desal. & Huge Desal. & Units  \\
        \hline
        $p_{capital}$ & Capital expense & 100 & 250 & 1000 & million $\S$/year \\
        $d_{capital}$ & Capital expense duration & 3 & 3 & 3 & year \\
        $p_{fixed}$ & Fixed expense & 1.0 & 2.5 & 10.0 & million $\S$/year \\
        $f_{water}^{currency}$ & Variable expense & 0.014 & 0.012 & 0.012 & $\S$/m$^3$ \\
        $q_{max}^{produce}$ & Maximum production & 50 & 150 & 600 & MCM/year \\
        $f_{water}^{elect}$ & Water-electricity intensity & 5.5 & 4.5 & 4.5 & kWh/m$^3$ \\
        \hline
        \multicolumn{6}{l}{Units: $\S$: fictional currency; MCM: million cubic meters; kWh: kilowatt hours}
    \end{tabular}
    \label{tab:wa_elements}
\end{table*}

The water system is coupled with the societal, agriculture, and electrical systems. Regional water demand arises from the societal and agriculture systems, $D_{water}(n) = D^{societal}_{water}(n) + D^{agricul}_{water}(n)$. Electricity resources required to satisfy the operational plan at each node total
\begin{equation}
    D^{water}_{elect}(n) = f^{elect}_{water}(n) \cdot q^{lift}_{water}(n) + \sum_{e\in\mathcal{E}(n)} f^{elect}_{water}(e) \cdot q^{produce}_{water}(e).
\end{equation}
At the end of each time step, the water system updates available aquifer stock based on withdrawals
\begin{equation}
    Q_{aquifer}(n, t+\Delta t) = Q_{aquifer}(n, t) - f^{aquifer}_{water}\cdot q^{lift}_{water}(n).
\end{equation}
The water system controller sets water production (desalination) levels in constituent infrastructure elements and determines the quantity of water to lift (from aquifers) and import at each node by solving the following linear program:

\noindent Find:
\begin{align}
    & q^{produce}_{water}(e) \; \forall \; e \in \mathcal{E} \\
    & q^{lift}_{water}(n), q_{water}^{import}(n) \; \forall \; n \in \mathcal{N}
\end{align}
Minimize:
\begin{align}
    \begin{split}
        & \sum_{e \in \mathcal{E}} \left( f^{currency}_{water}(e)+f^{elect}_{water}(e)\cdot \pi_{elect}^{local} \right)\cdot q^{produce}_{water}(e) \\
        &+ \sum_{n\in\mathcal{N}} \left( C \cdot q^{lift}_{water}(n) + \pi_{water}^{import} \cdot q_{water}^{import} (n) \right) \\
        & \text{where } \max_{e\in\mathcal{E}} \left( f^{currency}_{water}(e)+f^{elect}_{water}(e)\cdot \pi_{elect}^{local} \right) < C < \pi_{water}^{import}
    \end{split} 
\end{align}
Subject to:
\begin{align}
    & q^{produce}_{water}(e) \leq q^{produce}_{max}(e) \cdot b_{coastal}(n) \; \forall \; e \in \mathcal{E}(n) \; \forall \; n \in \mathcal{N} \\
    & \sum_{n\in\mathcal{N}} f_{water}^{aquifer}(n) \cdot q^{lift}_{water}(n) \leq Q_{aquifer}(n) \; \forall \; n \in \mathcal{N} \\
    & \sum_{e \in \mathcal{E}(n)} q^{produce}_{water}(e) + q_{water}^{import} (n) + q^{lift}_{water}(n) \geq D_{water}(n) \; \forall \; n \in \mathcal{N}
\end{align}

Net revenue for the water system includes revenue from domestic water production (lifting water is assumed to generate no direct revenue) and expenses from electricity to lift aquifer, import water, and capital, fixed, and variable costs based on lifecycle phase
\begin{equation}
    \begin{split}
        Q_{currency}^{water}(n) = & \pi^{local}_{water} \cdot \left( D_{water}(n) - q^{lift}_{water}(n) \right) \\
        & - \pi^{local}_{elect} \cdot f^{elect}_{water} \cdot q^{lift}_{water}(n) - \pi^{import}_{water} \cdot q^{import}_{water}(n) \\
        & - \sum_{e \in \mathcal{E}(n)}\left( p_{capital}(e) + p_{fixed}(e) + p_{variable}(e) \right)
    \end{split} 
\end{equation}
where variable costs include operating expenses and consumption of electricity resources
\begin{equation}
    p_{variable}(e) = \left( f_{water}^{currency}(e) + \pi^{local}_{elect}\cdot f_{water}^{elect}(e) \right) \cdot q^{produce}_{water}(e) .
\end{equation}

\subsection{Energy System Model}

Energy system properties in Table~\ref{tab:en_properties} define prices for domestic, imported, and exported oil resources and domestic electricity, set the initial stock of oil reservoirs in 1950, and set resources required to generate electricity to meet shortfalls. Only the industrial region has an oil reservoir and supply to the urban and rural regions must use pipelines.

\begin{table*}
    \footnotesize
    \caption{Energy system node properties.}
    \begin{tabular}{clcccl}
        \hline
        Variable & Description & Industrial & Urban & Rural & Units \\
        \hline
        $\pi^{local}_{oil}$ & Price of domestic oil (no net trade impact) & 8 & 8 & 8 & $\S$/toe \\
        $\pi^{import}_{oil}$ & Price of imported oil (net trade deficit) & 35 & 35 & 35 & $\S$/toe \\
        $\pi^{export}_{oil}$ & Price of exported oil (net trade deficit) & 30 & 30 & 30 & $\S$/toe \\
        $Q^{reservoir}_{0}$ & Initial reservoir volume & 65 & 0 & 0 & billion toe \\
        $\pi^{local}_{elect}$ & Price of electricity (no net trade impact) & 4 & 4 & 4 & $\S$/MWh \\
        $f^{oil}_{elect}$ & ``Private'' generation energy intensity & 0.5 & 0.5 & 0.5 & toe/MWh \\
        \hline
        \multicolumn{6}{l}{Units: $\S$: fictional currency; toe: ton of oil equivalent; MWh: megawatt hour}
    \end{tabular}
    \label{tab:en_properties}
\end{table*}

Petroleum element properties in Table~\ref{tab:pe_elements} define two sizes of wells and two sizes of pipelines. Although oil refining typically includes numerous feed stock types and output products, this model assumes wells directly produce consumable oil at a one-to-one ratio from the reported reservoir stock. Despite large capital and operations expenses of associated infrastructure, oil production is very profitable due to high export prices.

\begin{table*}
    \footnotesize
    \caption{Petroleum system element properties.}
    \begin{tabular}{clccccl}
        \hline
        Variable & Description & Sm.~Well & Lg.~Well & Sm.~Pipe & Lg.~Pipe & Units  \\
        \hline
        $p_{capital}$ & Capital expense & 500 & 875 & 100 & 300 & million $\S$/year \\
        $d_{capital}$ & Capital expense duration & 3 & 3 & 3 & 3 & year \\
        $p_{fixed}$ & Fixed expense & 25.0 & 87.5 & 2.0 & 9.0 & million $\S$/year \\
        $f_{oil}^{currency}$ & Variable expense & 6.00 & 5.75 & 0.10 & 0.10 & $\S$/toe \\
        $q_{max}^{produce}$ & Maximum production & 25 & 100 & -- & -- & million toe/year \\
        $f_{oil}^{reservoir}$ & Oil-reservoir intensity & 1.0 & 1.0 & -- & -- & toe/toe \\
        $f_{oil}^{elect}$ & Oil-electricity intensity & -- & -- & 2 & 2 & kWh/toe \\
        $q_{max}^{transport}$ & Maximum throughput & -- & -- & 10 & 50 & million toe/year \\
        $\eta$ & Transport efficiency & -- & -- & 98 & 99 & \% \\
        \hline
        \multicolumn{7}{l}{Units: $\S$: fictional currency; toe: ton of oil equivalent; kWh: kilowatt hour}
    \end{tabular}
    \label{tab:pe_elements}
\end{table*}

Electrical element properties in Table~\ref{tab:el_elements} define two sizes of plants for thermal and renewable generation based on solar photo-voltaic technology. No distribution elements are available to transport electricity between regions. Thermal generation consumes oil as the primary operational cost to create electricity and is up to twice as efficient as the default method used to satisfy insufficient supply. Solar generation has no variable operating expenses but incurs a large initial capital expense and larger fixed operations expenses compared to thermal generation.

\begin{table*}
    \footnotesize
    \caption{Electrical system element properties.}
    \begin{tabular}{clccccl}
        \hline
        Variable & Description & S.~Thermal & L.~Thermal & S.~Solar & L.~Solar & Units  \\
        \hline
        $p_{capital}$ & Capital expense & 25 & 75 & 100 & 450 & million $\S$/year \\
        $d_{capital}$ & Capital expense duration & 2 & 3 & 3 & 3 & year \\
        $p_{fixed}$ & Fixed expense & 0.25 & 1.50 & 3.00 & 13.50 & million $\S$/year \\
        $f_{elect}^{currency}$ & Variable expense & 0 & 0 & 0 & 0 & $\S$/MWh \\
        $q_{max}^{produce}$ & Maximum production & 2 & 10 & 2 & 10 & TWh/year \\
        $f_{elect}^{oil}$ & Electricity-oil intensity & 0.30 & 0.25 & 0 & 0 & toe/MWh \\
        \hline
        \multicolumn{7}{l}{Units: $\S$: fictional currency; toe: ton of oil equivalent; MWh: megawatt hour; TWh: terawatt hour}
    \end{tabular}
    \label{tab:el_elements}
\end{table*}

In addition to the internal mutual dependency, the energy system is coupled with societal and water systems. Regional oil demand arises from societal and electricity systems, $D_{oil}(n) = D^{societal}_{oil}(n) + D^{elect}_{oil}(n)$ and regional electricity demand arises from societal, water, and oil systems, $D_{elect}(n) = D^{societal}_{elect}(n) + D^{water}_{elect}(n) + D^{oil}_{elect}(n)$. Electricity resources required to satisfy petroleum system operations at each node total
\begin{equation}
    D^{petrol}_{elect}(n) = \sum_{e\in\mathcal{E}(n)} f^{elect}_{oil}(e) \cdot q^{transport}_{oil}(e).
\end{equation}
Oil resources required to satisfy electricity system operations at each node total
\begin{equation}
    D^{elect}_{oil}(n) = f^{oil}_{elect}(n) \cdot q^{produce}_{elect}(n) + \sum_{e\in\mathcal{E}(n)} f^{oil}_{elect}(e) \cdot q^{produce}_{elect}(e).
\end{equation}
At the end of each time step, the petroleum system updates available reservoir stock based on withdrawals
\begin{equation}
    Q_{reservoir}(n, t+\Delta t) = Q_{reservoir}(n, t) - \sum_{e\in\mathcal{E}(n)} f^{reservoir}_{oil}(e)\cdot q^{produce}_{oil}(e).
\end{equation}
Net energy system revenue includes petroleum and electricity sources: $Q_{currency}^{energy}(n)=q_{currency}^{petrol}(n)+q_{currency}^{elect}(n)$.

The petroleum system controller sets oil production (from reservoirs) and transport levels in constituent infrastructure elements and determines the quantity of oil to import and export at each node by solving the following linear program:

\noindent Find:
\begin{align}
    & q^{produce}_{oil}(e), q^{transport}_{oil}(e) \; \forall \; e \in \mathcal{E} \\
    & q_{oil}^{import}(n), q_{oil}^{export}(n) \; \forall \; n \in \mathcal{N}
\end{align}
Minimize:
\begin{align}
    \begin{split}
        &\sum_{e \in \mathcal{E}} f^{currency}_{oil}(e) \cdot q^{produce}_{oil}(e) \\
        &+ \sum_{e \in \mathcal{E}} \left( f^{currency}_{oil}(e)+f^{elect}_{oil}(e)\cdot \pi_{elect}^{local} \right)\cdot q^{transport}_{oil}(e) \\
        &+ \sum_{n\in\mathcal{N}} \left( \pi_{oil}^{import} \cdot q_{oil}^{import} (n) - \pi_{oil}^{export} \cdot q_{oil}^{export} (n)\right) \\
    \end{split} 
\end{align}
Subject to:
\begin{align}
    & q^{produce}_{oil}(e) \leq q^{produce}_{max}(e) \; \forall \; e \in \mathcal{E} \\
    & q^{transport}_{oil}(e) \leq q^{transport}_{max}(e) \; \forall \; e \in \mathcal{E} \\
    & \sum_{e \in \mathcal{E}(n)} f_{oil}^{reservoir}(e) \cdot q^{produce}_{oil}(e) \leq Q_{reservoir}(n) \; \forall \; n \in \mathcal{N} \\
    \begin{split}
        & \sum_{e \in \mathcal{E}(n)} \left( q^{produce}_{oil}(e) - q^{transport}_{oil}(e) \right) + \sum_{e \in \mathcal{E}^\prime(n)} \eta(e)\cdot q^{transport}_{oil}(e) \\
        & + q_{oil}^{import} (n) - q_{oil}^{export} (n) \geq D_{oil}(n) \; \forall \; n \in \mathcal{N}
    \end{split}
\end{align}

Petroleum system net revenue includes revenue from local, regional distribution, and export sales, resource expenses from regional distribution and import purchases, and other expenses from capital, fixed, and/or variable costs based on lifecycle phase
\begin{equation}
    \begin{split}
        q_{currency}^{petrol}(n) = & \pi^{local}_{oil} \cdot D_{oil}(n) + \sum_{e \in \mathcal{E}(n)} \left( \pi^{local}_{oil} \cdot \eta(e) \cdot q^{transport}_{oil}(e) \right) + \pi^{export}_{oil} \cdot q^{export}_{oil}(n) \\
        & - \sum_{e \in \mathcal{E}^\prime(n)} \left( \pi^{local}_{oil} \cdot \eta(e) \cdot q^{transport}_{oil}(e) \right) - \pi^{import}_{oil} \cdot q^{import}_{oil}(n) \\
        & - \sum_{e \in \mathcal{E}(n)}\left( p_{capital}(e) + p_{fixed}(e) + p_{variable}(e) \right)
    \end{split} 
\end{equation}
where variable costs include operating expenses and consumption of electricity resources
\begin{equation}
    p_{variable}(e) = \left( f_{oil}^{currency}(e) + \pi^{local}_{elect}\cdot f_{oil}^{elect}(e) \right) \cdot \left( q^{produce}_{oil}(e) + q^{transport}_{oil}(e) \right).
\end{equation} 

The electricity system controller sets electricity generation levels in constituent infrastructure elements and determines the quantity of electricity to generate from low-efficiency methods at each node by solving the following linear program:

\noindent Find:
\begin{align}
    & q^{produce}_{elect}(e) \; \forall \; e \in \mathcal{E} \\
    & q^{private}_{elect}(n) \; \forall \; n \in \mathcal{N}
\end{align}
Minimize:
\begin{align}
    \begin{split}
        &\sum_{e \in \mathcal{E}} \left( f^{currency}_{elect}(e) + f^{oil}_{elect}(e) \cdot \pi^{local}_{oil} \right) \cdot q^{produce}_{elect}(e) + \sum_{n\in\mathcal{N}} C \cdot q^{produce}_{elect}(n) \\
        & \text{where } C > \max_{e\in\mathcal{E}} \left( f^{currency}_{elect}(e)+f^{oil}_{elect}(e)\cdot \pi_{oil}^{local} \right)
    \end{split} 
\end{align}
Subject to:
\begin{align}
    & q^{produce}_{elect}(e) \leq q^{produce}_{max}(e) \; \forall \; e \in \mathcal{E} \\
    \begin{split}
        & \sum_{e \in \mathcal{E}(n)} q^{produce}_{elect}(e) + q^{private}_{elect}(n) \geq D_{elect}(n) \; \forall \; n \in \mathcal{N}
    \end{split}
\end{align}

Electricity system net revenue includes revenue from domestic generation (private generation is assumed to generate no direct revenue) and expenses from oil for private generation and capital, fixed, and variable costs based on lifecycle phase
\begin{equation}
    \begin{split}
        q_{currency}^{elect}(n) = & \pi^{local}_{elect} \cdot \left( D_{elect}(n) - q^{private}_{elect}(n) \right) - \pi^{local}_{oil}\cdot f^{oil}_{elect} \cdot q^{private}_{elect}(n) \\
        & - \sum_{e \in \mathcal{E}(n)}\left( p_{capital}(e) + p_{fixed}(e) + p_{variable}(e) \right)
    \end{split} 
\end{equation}
where variable costs include operating expenses and consumption of electricity resources
\begin{equation}
    p_{variable}(e) = \left( f_{elect}^{currency}(e) + \pi^{local}_{oil}\cdot f_{elect}^{oil}(e) \right) \cdot q^{produce}_{elect}(e) .
\end{equation}

\subsection{Key Model Assumptions and Limitations}

The driving forces from population and societal resource demand dynamics are fixed and exogenous to the model formulation. A more realistic population growth model would link growth rates to a measure of economic performance or environmental state to simulate the consequences of depleted resources or comparative prosperity of economic booms.

All resource prices are static, homogeneous across regions, and exogeneous from the model formulation. Price points are approximately based on marginal costs of production. A more realistic (but much more complex) resource pricing model would establish market conditions based on supply capacity and demand to determine equilibrium price conditions at each time step where variation across regions could generate new pressures on infrastructure.

Available infrastructure projects include a fixed set of elements with static properties. While some properties are based on current technology and physical limits of transformation, others are fit to establish internal consistency (e.g.~return on investment periods). A more detailed model would allow variable capacities with economies of scale and efficiency improvements or new technology options over time. For example, the past 30 years have observed tremendous improvements in renewable power generation technologies and agriculture yields.

The model assumes a centrally-managed ``nationalized'' infrastructure perspective that is an over-simplification of any economy. For example, the agriculture system is the sole source of domestic food and subsidizes imported food at the local price. Two exceptions include lifted water from aquifers and private electricity generation which both provide resources without infrastructure but do incur expenses from resource consumption (electricity and oil, respectively).

Finally, the model assumes deterministic dynamic behavior aggregated to annual periods to mitigate logistical effects of delays and buffers. This assumes demands to be satisfied at some point during the year-long period, ignoring seasonal variation, and that constituent infrastructure can be operated efficiently without surplus resources that must be discarded. While a finer timescale and stochastic features exemplify real-world planning complexity, considering them would exceed available time and cognitive bandwidth for co-design sessions.

\section{Objective Metric Formulation} \label{sec:ob_details}

This section provides details about the role-specific and joint objectives. Most objectives are expressed as a time average over the 30-year planning period to mitigate boundary effects.

\subsection{Food Security}

Food security measures the average fraction of domestic food supply between 1980 and $t>1980$ compared to a desired value of 75\%. It ranges between 0 for no domestic food production in all years to 1000 for at least 75\% domestic food production in all years. It is computed for year $t$ as:
\begin{equation}
    J_{food}(t) = \frac{1000}{t-1980}\sum_{i=1980}^t F(i)
\end{equation}
where
\begin{align}
    F(i) &= \begin{cases}
        1 & \text{if } S(i)/D(i) \geq 0.75 \\
        0 & \text{if } S(i)/D(i) < 0 \\
        \frac{S(i)/D(i)}{0.75} & \text{otherwise}
    \end{cases}\\
    S(i) &= \sum_{e\in\mathcal{E}} f^{land}_{food}(e,i)\cdot q^{use}_{land}(e,i)\\
    D(i) &= \sum_{n\in\mathcal{N}} D_{food}(n,i)
\end{align}

\subsection{Aquifer Security}

Aquifer security measures the average expected lifetime of an aquifer between 1980 and $t>1980$ compared to a desired value of 200 years. It ranges between 0 for an expected lifetime less than 20 years in all years to 1000 if above 200 years in all years. It is computed for year $t$ as:
\begin{equation}
    J_{aquifer} = \frac{1000}{t-1980} \sum_{i=1980}^t L_{a}(i)
\end{equation}
where
\begin{align}
    L_{a}(i) &= \begin{cases}
        1 & \text{if } V_{a}(i)/W_{a}(i) \geq 200 \\
        0 & \text{if } V_{a}(i)/W_{a}(i) < 20 \\
        \frac{V_{a}(i)/W_{a}(i) - 20}{200 - 20} & \text{otherwise}
    \end{cases}\\
    V_{a}(i) &= \sum_{n\in\mathcal{N}} Q_{aquifer}(n,i) \\
    W_{a}(i) &= \sum_{n\in\mathcal{N}} f^{aquifer}_{water}(n,i)\cdot q^{lift}_{water}(n,i)
\end{align}

\subsection{B.3. Reservoir Security}

Reservoir security measures the average expected lifetime of an oil reservoir between 1980 and $t>1980$ compared to a desired value of 200 years. It ranges between 0 for no remaining lifetime in all years to 1000 for an expected lifetime above 200 years in all years. It is computed for year $t$ as:
\begin{equation}
    J_{reservoir} = \frac{1000}{t-1980} \sum_{i=1980}^t L_{r}(i)
\end{equation}
where
\begin{align}
    L_{r}(i) &= \begin{cases}
        1 & \text{if } V_{r}(i)/W_{r}(i) \geq 200 \\
        0 & \text{if } V_{r}(i)/W_{r}(i) < 0 \\
        \frac{V_{r}(i)/W_{r}(i)}{200} & \text{otherwise}
    \end{cases}\\
    V_{r}(i) &= \sum_{n\in\mathcal{N}} Q_{r}(n,i) \\
    W_{r}(i) &= \sum_{e\in\mathcal{E}} f^{r}_{oil}(e,i)\cdot q^{produce}_{oil}(e,i)
\end{align}

\subsection{Financial Security}

Financial security measures the cumulative net revenue earned compared to a minimum and maximum desired values. It represents motivation of a player to operate profitable infrastructure and ranges between 0 if the lower bound is not achieved in all years and 1000 if the upper bound is achieved in all years. It is computed for year $t$ as:
\begin{equation}
    J_{financial} = \begin{cases}
        1000 & \text{if } R(t) > R_{max}(t) \\
        0 & \text{if } R(t) < R_{min}(t) \\
        \frac{R(t)-R_{min}(t)}{R_{max}(t)-R_{min}(t)} & \text{otherwise}
    \end{cases}\\
\end{equation}
where
\begin{align}
    R(t) &= \sum_{i=1980}^t \sum_{n\in\mathcal{N}} Q_{currency}^{sector}(n,i) \\
    R_{min}(t) &= R_{min}^{2010} \cdot \frac{(1+r_R)^{t-1940} - 1}{(1+r_R)^{2010-1940} - 1} \\
    R_{max}(t) &= R_{max}^{2010} \cdot \frac{(1+r_R)^{t-1940} - 1}{(1+r_R)^{2010-1940} - 1}
\end{align}
using sector-specific model parameters $R_{min}^{2010}$, $R_{max}^{2010}$, and $r_R$ in Table~\ref{tab:objective_parameters}.

\begin{table}
    \centering
    \caption{Financial Security and Political Power Objective Function Parameters}
    \begin{tabular}{ccccccc}
        \hline
        Sector & \multicolumn{3}{c}{Financial Security} & \multicolumn{2}{c}{Political Power} \\
        & $R_{min}^{2010}$ & $R_{max}^{2010}$ & $r_R$ & $I_{2010}$ & $r_I$ \\
        \hline
        Agriculture & 0 & $\S$50 billion & 5\% & $\S$10 billion & 6\% \\
        Water & $-\S$10 billion & 0 & 6\% & $\S$15 billion & 6\% \\
        Energy & 0 & $\S$500 billion & 4\% & $\S$50 billion & 3\% \\
        Joint/Shared & $-\S$10 billion & $\S$550 billion & 4\% & -- & --\\
        \hline
    \end{tabular}
    \label{tab:objective_parameters}
\end{table}

\subsection{Political Power}

Political power measures the cumulative capital expenses allocated to a sector compared to a minimum desired value. It represents the motivation of a player to acquire funds from a limited national budget and ranges between 0 if there is no cumulative capital investment up to year $t$ and 1000 if the cumulative capital investment exceeds an upper bound. It is computed for year $t$ as
\begin{equation}
    J_{political} = \begin{cases}
        1000 & \text{if } I(t) > I_{max}(t) \\
        \frac{I(t)}{I_{max}(t)} & \text{otherwise}
    \end{cases}\\
\end{equation}
where
\begin{align}
    I(t) &= \sum_{i=1980}^t \sum_{e\in\mathcal{E}} p_{capital}(e,i) \\
    I_{max}(t) &= I_{2010} \cdot  \frac{(1+r_I)^{t-1940} - 1}{(1+r_I)^{2010-1940} - 1}
\end{align}
using sector-specific model parameters $I_{2010}$ and $r_I$ in Table~\ref{tab:objective_parameters}.

\bibliographystyle{apalike}
\bibliography{main}

\end{document}